\documentclass[11pt]{amsart}
\usepackage{graphicx} 
\usepackage{amsmath,amsfonts,amsthm,amssymb,amscd}
\usepackage{sansmath}
\usepackage{lipsum}
\usepackage{xcolor}
\usepackage{amsfonts}
\usepackage{graphicx}
\usepackage{epstopdf}
\usepackage{algorithm}
\usepackage{algorithmic}
\usepackage{upgreek}
\usepackage{overpic}
\usepackage{enumerate}
\usepackage{enumitem}
\usepackage{bm}
\usepackage[hidelinks]{hyperref}

\usepackage{booktabs}
\hypersetup{
  colorlinks=true,
  linkcolor=black,     
  citecolor=blue,      
  urlcolor=black,      
  filecolor=black      
}
\usepackage[foot]{amsaddr}
\usepackage[letterpaper,top=3.5cm,bottom=3.5cm,left=3.5cm,right=3.5cm,marginparwidth=1.75cm]{geometry}

\newcommand{\cN}{\mathcal{N}}

\newcommand{\cD}{\mathcal{D}}

\newcommand{\bE}{\mathbb{E}}

\newcommand{\bR}{\mathbb{R}}

\newcommand{\bN}{\mathbb{N}}

\newcommand{\vx}{\mathbf{x}}
\newcommand{\vy}{\mathbf{y}}
\newcommand{\vm}{\mathbf{m}}

\newcommand{\vp}{\mathbf{p}}
\newcommand{\vr}{\mathbf{r}}
\newcommand{\vq}{\mathbf{q}}

\newcommand{\sfL}{\mathsf{L}}
\newcommand{\sfP}{\mathsf{P}}

\newtheorem{theorem}{Theorem}[section]

\newtheorem{proposition}[theorem]{Proposition}

\theoremstyle{remark}
\newtheorem{remark}[theorem]{Remark}
\numberwithin{equation}{section}
\newenvironment{newremark}[1]{%
    \begin{remark}#1}{%
    \Endofdef\end{remark}%
}
\newcommand{\xqed}[1]{%
    \leavevmode\unskip\penalty9999 \hbox{}\nobreak\hfill
    \quad\hbox{\ensuremath{#1}}}
\newcommand{\Endofdef}{\xqed{\lozenge}}

\title[Affine invariant ensemble samplers and their dimensional scaling]{New affine invariant ensemble samplers and their dimensional scaling}

\author{Yifan Chen}
\address{Department of Mathematics, University of California, Los Angeles, CA, USA}
\email{yifanchen@math.ucla.edu}
\date{}

\begin{document}

\begin{abstract}
We introduce new affine invariant ensemble Markov chain Monte Carlo (MCMC) samplers that are easy to construct and improve upon existing methods, especially for high-dimensional problems. We first propose a simple derivative-free side move sampler that improves upon popular samplers in the \texttt{emcee} package by generating more effective proposal directions. We then develop a class of derivative-based affine invariant ensemble Hamiltonian Monte Carlo (HMC) samplers based on antisymmetric preconditioning using complementary ensembles, which outperform standard, non-affine-invariant HMC when sampling highly anisotropic distributions.
We provide asymptotic scaling
analysis for high-dimensional Gaussian targets to further elucidate the properties of
these affine invariant ensemble samplers. In particular, with derivative information,
the affine invariant ensemble HMC can scale much better with dimension compared
to derivative-free ensemble samplers.

\end{abstract}

\maketitle
\vspace{-2em}
\tableofcontents
\vspace{-1em}
\section{Introduction}
The concept of affine invariance \cite{goodman2010ensemble} has played an important role in the development of efficient Markov chain Monte Carlo (MCMC) samplers as it provides robustness against anisotropy and obviates the need for problem-specific preconditioning. Samplers are called affine invariant if they behave consistently across all coordinate systems related through affine transformations. In practice, affine invariance in MCMC is commonly achieved through ensemble samplers, which propose updates based on relative configurations of multiple interacting chains.

Affine invariant ensemble samplers have proven effective for routine Bayesian inference applications \cite{foreman2013emcee}, but they may perform poorly in high-dimensional problems \cite{huijser2015properties, carpenter2017ensemble}. This particularly applies to derivative-free samplers such as the stretch move sampler \cite{goodman2010ensemble}, widely used via the \texttt{emcee} package \cite{foreman2013emcee}.
Indeed, compelling arguments suggest that ``ensemble methods are doomed to fail in high dimensions'' \cite{carpenter2017ensemble}, based on insightful observations about typical sets: high-dimensional distributions such as Gaussians typically concentrate on thin shells, and the interpolation or extrapolation between points in the stretch move---as well as in many other ensemble samplers---is unlikely to fall within this shell. As a result, small step sizes are required to maintain reasonable acceptance rates, and the samplers effectively ``devolve into random walks with poorly biased directional choices.''

This paper revisits these concerns and shows that the situation is more nuanced. We introduce new affine invariant ensemble samplers and analyze their scaling behavior with dimension. We demonstrate that ensemble methods can, in fact, achieve the same high-dimensional scaling as their single-chain counterparts, both in the derivative-free and derivative-based cases and at the stationary phase. In particular, ensemble samplers that incorporate gradient information can overcome random-walk behavior and scale more favorably with dimension. Moreover, due to affine invariance, the constants hidden in the asymptotic scaling are insensitive to the condition number of the target distribution.

\subsection{This work}
We summarize our contributions as follows.
\subsubsection{Derivative-free ensemble samplers}
We first propose a derivative-free \emph{side move} ensemble sampler that improves upon the stretch move by adopting a more favorable proposal direction in high dimensions. For Gaussian targets, we analyze the acceptance probability and show that the optimal step-size parameters—denoted $a-1$ for the stretch move and $\sigma$ for the side move—scale as $d^{-1/2}$. This scaling matches that of the single-chain random walk Metropolis algorithm \cite{gelman1997weak, yang2020optimal}. We further show that, under optimal scaling, the expected squared jumping distance \cite{pasarica2010adaptively, atchade2011towards, roberts2014minimising, yang2020optimal} is larger for the side move than for the stretch move.

Our numerical experiments confirm these theoretical predictions: the integrated autocorrelation times of both samplers scale linearly with the dimension $d$, while the side move consistently outperforms the stretch move—often by a factor of two or more—across a range of problems in dimensions on the order of $100$.

\subsubsection{Derivative-based ensemble samplers}
Beyond derivative-free methods, it is well known that incorporating gradient information allows MCMC samplers to scale more favorably with dimension \cite{roberts1998optimal, beskos2013optimal}. In particular, Hamiltonian Monte Carlo (HMC) \cite{neal2011mcmc} avoids random-walk behavior and achieves the state-of-the-art $d^{-1/4}$ scaling for the step size.

Whether one can construct an efficient {affine invariant} ensemble HMC sampler has remained an open question. In this work, we propose a new class of affine invariant ensemble HMC methods, one of which attains $d^{-1/4}$ step-size scaling for high-dimensional Gaussian targets. The key idea is to employ an antisymmetric preconditioning strategy derived from complementary ensembles, enabling efficient parallel updates without relying on traditional mass-matrix tuning as in standard HMC.

Our experiments demonstrate that these affine invariant HMC samplers can outperform derivative-free ensemble methods by an order of magnitude and achieve 10- to 100-fold reductions in autocorrelation time compared to standard HMC (without elaborate tuning of mass matrices) when sampling from $\sim$100-dimensional synthetic anisotropic distributions and from distributions arising in stochastic PDE models, which are typically ill-conditioned.

In summary, while ensemble samplers based purely on derivative-free interpolation or extrapolation inevitably exhibit random-walk behavior and struggle in high dimensions, ensemble samplers that incorporate derivative information in the spirit of HMC avoid this limitation and are therefore not doomed to fail in high-dimensional settings.

\subsection{Related work on affine invariant samplers} 
The concept of affine invariance, introduced to MCMC samplers by \cite{goodman2010ensemble}, draws inspiration from the empirical success of the Nelder-Mead simplex algorithm \cite{nelder1965simplex} in optimization. This idea has been extended to many other sampling and related areas, including data assimilation \cite{reich2015probabilistic}, annealed importance sampling \cite{chen2024ensemble}, and variational inference \cite{chen2023sampling}, to list a few.

The stretch move \cite{goodman2010ensemble}, which is related to the scaling-invariant $t$-walk move proposed in \cite{christen2007general}, remains one of the most widely used affine invariant ensemble samplers. Popularized through the \texttt{emcee} package \cite{foreman2013emcee}, it has been extensively adopted in the astrophysics community. The stretch move is most effective in moderate dimensions (e.g., $d \leq 20$) but must be used cautiously for high-dimensional problems \cite{huijser2015properties}. To address high-dimensional or infinite-dimensional problems, particularly in PDEs and inverse problems, researchers have developed hybrid approaches that combine function space MCMC \cite{cotter2013mcmc} with ensemble samplers \cite{coullon2021ensemble, dunlop2022gradient}.

The stretch move is closely related to the snooker algorithm \cite{gilks1994adaptive, roberts1994convergence}, which uses ensembles to enable adaptive directional sampling. The ensemble covariance has also been used to design derivative-free affine invariant samplers, such as the walk move \cite{goodman2010ensemble}. Another well-known ensemble sampler is differential evolution MCMC \cite{braak2006markov, vrugt2009accelerating}, which is inspired by the differential evolution optimization algorithm \cite{storn1997differential}. The affine invariant side move we introduce in this paper is connected to both the walk move and differential evolution MCMC, as we detail in Section \ref{sec-side-move-sampler}.

Beyond derivative-free samplers, derivative-based affine invariant samplers have been developed, including those based on Riemannian geometry \cite{girolami2011riemann} and Newton-type directions \cite{martin2012stochastic, simsekli2016stochastic, detommaso2018stein, chewi2020exponential}. Affine invariance can also be achieved by combining ensemble covariance preconditioning with gradient-based directions \cite{greengard2015ensemblized, leimkuhler2018ensemble, garbuno2020interacting, garbuno2020affine, liu2022second, chen2023sampling}. These methods establish mathematical links to gradient flows and ensemble Kalman filters via derivative-free approximations \cite{garbuno2020interacting, garbuno2020affine, chen2023sampling}. However, most of these approaches do not incorporate Metropolis adjustments and are therefore biased. Our focus in this paper is on adjusted, unbiased MCMC samplers.

Among unbiased derivative-based MCMC samplers, HMC \cite{neal2011mcmc} achieves the state-of-the-art scaling with dimension \cite{beskos2013optimal}. However, HMC can struggle with anisotropic or ill-conditioned target distributions and requires careful adaptation of the mass matrix \cite{carpenter2017stan} to correct scale imbalances, as it is not affine invariant. Metrics based on Riemannian manifolds and Hessians have been used to make HMC affine invariant \cite{girolami2011riemann, lee2018convergence, kook2022sampling}, but these approaches generally require costly second-order information. We believe that systematically developing affine invariant ensemble HMC methods — based solely on first-order information — represents a promising algorithmic opportunity to advance the state of ensemble samplers.
Related work has explored diagonal preconditioning based on ensemble variance \cite{hoffman2022tuning}, yielding scale invariance—a weaker notion than affine invariance—as well as tuning-free multichain adaptive HMC methods designed for parallel architectures \cite{hoffman2021adaptive, sountsov2021focusing, sountsov2024running}.

\subsection{Organization} 
We review the affine invariance concept in Section \ref{sec-affine-invariance}. We introduce the side move sampler in Section \ref{sec-side-move-sampler} and provide scaling analysis in Section \ref{sec-Formal analysis of high dimensional behaviors}. Section \ref{sec-Affine Invariant Hamiltonian Monte Carlo Samplers} discusses the affine invariant HMC sampler, followed by scaling analysis in Section \ref{sec-Analysis of high dimensional scaling}. We present numerical experiments in Section \ref{sec-Numerical Experiments} and conclude in Section \ref{sec-Discussions and Conclusions}. Appendix \ref{appendix-Pseudocode of the Algorithms} includes pseudocode for all algorithms, while Appendices \ref{Proof-scaling-side-stretch} and \ref{Proof-scaling-HMC} contain technical proofs.

\section{Affine Invariance}
\label{sec-affine-invariance}
We review the mathematical concept of affine invariance. Following the seminal work of Goodman and Weare \cite{goodman2010ensemble}, we express a general MCMC sampler as
\[\mathbf{x}(m+1) = R(\mathbf{x}(m), \pi)\]
where $R$ denotes a mapping at iteration $m \in \mathbb{N}$ that is random. This randomness typically depends on the random numbers generated during the proposal step and in the accept-reject procedure. The sampler is affine invariant if, for any invertible affine transformation $\mathbf{y} = \phi(\mathbf{x}) = A\mathbf{x} + \mathbf{b}$, the following equality holds for a given realization of the mapping $R$ (i.e., by fixing the random number seeds):
\begin{equation}
\label{eqn-affine-invariant}
    \mathbf{y}(m+1) = R(\mathbf{y}(m), \phi\#\pi) \, .
\end{equation}
Here, $\phi\#\pi$ denotes the push-forward density of $\pi$ under $\phi$, defined by
\[(\phi\#\pi)(\mathbf{y}) = |\det A|^{-1} \pi(\phi^{-1}(\mathbf{y}))\, .\]
Equation \eqref{eqn-affine-invariant} indicates that the transformed sequence $\{\mathbf{y}(m)\}$ coincides with what would be obtained by applying the same MCMC algorithm to the transformed density $\phi\#\pi$ with initial value $\mathbf{y}(0)$. Consequently, the convergence behavior of $\mathbf{x}(m)$ toward sampling $\pi$ matches that of $\mathbf{y}(m)$ toward sampling $\phi\#\pi$. In particular, the sampler inherits the convergence efficiency of the optimally preconditioned coordinate system that minimizes anisotropy through affine transformation.

Affine invariance in MCMC has been widely achieved through ensemble samplers. The concept extends naturally to the ensemble setting. We denote the positions of an ensemble of particles, or walkers, by $(\mathbf{x}_1(m),\mathbf{x}_2(m),\ldots,\mathbf{x}_N(m))$ at discrete time step $m \in \mathbb{N}$, with $m=0$ representing the initial configuration. We can view the ensemble sampler as operating in the extended space $\mathbb{R}^{dN}$, targeting the product distribution $\pi^N$, which is the product of $N$ independent copies of $\pi$. Similarly, we denote a general ensemble MCMC sampler as
\[(\mathbf{x}_1(m+1),\mathbf{x}_2(m+1),\ldots,\mathbf{x}_N(m+1)) = R(\mathbf{x}_1(m),\mathbf{x}_2(m),\ldots,\mathbf{x}_N(m), \pi)\, .\]
This sampler is affine invariant if, for any invertible affine transformation $\mathbf{y} = \phi(\mathbf{x}) = A\mathbf{x}+\mathbf{b}$ such that
\begin{equation*}
    (\mathbf{x}_1,\ldots,\mathbf{x}_N) \overset{\phi}{\to} (\mathbf{y}_1,\ldots,\mathbf{y}_N) = (A\mathbf{x}_1+\mathbf{b},\ldots,A\mathbf{x}_N+\mathbf{b})\, ,
\end{equation*}
it holds that
\[(\mathbf{y}_1(m+1),\mathbf{y}_2(m+1),\ldots,\mathbf{y}_N(m+1)) = R(\mathbf{y}_1(m),\mathbf{y}_2(m),\ldots,\mathbf{y}_N(m), \phi\#\pi)\, .\]
\section{Derivative-free Side Move Sampler}
\label{sec-side-move-sampler}
    \subsection{Basic side move} We denote the positions of an ensemble at discrete time step $m \in \mathbb{N}$ as $(\mathbf{x}_1(m),\mathbf{x}_2(m),...,\mathbf{x}_N(m))$, with $m=0$ representing the initial configuration. We require $N > d$ and that the $N$ particles span the full space $\mathbb{R}^d$.


At each time step, our ensemble \textit{side move} sampler randomly selects one particle $\mathbf{x}_i(m)$ and two distinct particles $\mathbf{x}_j(m)$ and $\mathbf{x}_k(m)$ from the ensemble which are different from $\mathbf{x}_i(m)$. The sampler proposes the following \textit{side move} (see an illustration in Figure \ref{fig:stretch-side-moves}) for the $i$-th particle:
\begin{equation}
    \tilde{\vx}_i(m+1) = \vx_i(m) + \sigma (\vx_j(m)-\vx_k(m)) \xi\, ,
\end{equation}
where $\sigma$ is a user-specified scalar parameter, and $\xi \sim \mathcal{N}(0,1)$ is drawn from a normal distribution. We will discuss connections to walk move \cite{goodman2010ensemble} and differential evolution \cite{storn1997differential,braak2006markov, vrugt2009accelerating} in Sections  \ref{sec-Comparison to other affine invariant moves} and \ref{sec-Connection to differential evolution}. In Section \ref{sec-Formal analysis of high dimensional behaviors}, we study the high dimensional scaling behavior, which suggests a choice of $\sigma = 1.687d^{-1/2}$.

The above proposal for the $i$-th particle is accepted or rejected according to the standard Metropolis criterion. Since for fixed $j$ and $k$, the proposal is symmetric for the $i$-th particle, we get a simple acceptance probability:
\begin{equation}
    \mathrm{prob} = \min\, \{1, \frac{\pi(\tilde{\vx}_i(m+1))}{\pi(\vx_i(m))} \}\, .
\end{equation}
With this probability, we update
    $\vx_i(m+1) = \tilde{\vx}_i(m+1)$,
otherwise the proposal is rejected and we set $\vx_i(m+1) = \vx_i(m)$. All other particles (those other than the $i$-th) remain unchanged during this step.

The update preserves $\pi^N$ as an invariant distribution. In fact, this update can be viewed as a Metropolis-within-Gibbs approach to sample from $\pi^N$: at each step, we propose an update that satisfies detailed balance for the conditional distribution of one particular particle, given the positions of all other particles in the ensemble.

The affine invariance property of the algorithm is apparent since it is based on relative positions of the particles. We include a derivation here for completeness. 
Given an invertible affine transformation $\phi(\mathbf{x}) = A\mathbf{x} + \mathbf{b}$, we transform the initial ensemble
\begin{equation}
    (\vx_1,...,\vx_N) \overset{\phi}{\to} (\vy_1,...,\vy_N) = (A\vx_1+\mathbf{b},...,A\vx_N+\mathbf{b})\, .
\end{equation}
Using the side move algorithm to sample $\phi\# \pi$ with the transformed initial ensemble, we get the proposal (assuming the same random number generator and seeds)
\[\tilde{\vy}_i(m+1) = \vy_i(m) + \sigma (\vy_j(m)-\vy_k(m)) \xi = A\tilde{\vx}_i(m+1) + \mathbf{b}\, . \]
Moreover, by the change of variables, the acceptance ratio remains the same as the untransformed case:
\[ \frac{\pi(\tilde{\vx}_i(m+1))}{\pi(\vx_i(m))}  = \frac{(\phi\#\pi)(\tilde{\vy}_i(m+1))}{(\phi\#\pi)(\vy_i(m))} \, . \]
Therefore, we obtain that for any $m \in \bN$, it holds
\[(\vy_1(m),...,\vy_N(m)) = (A\vx_1(m)+\mathbf{b},...,A\vx_N(m)+\mathbf{b})\, , \]
which justifies the affine invariance property.
\subsection{Parallel side move sampler}
\label{sec-parallel-side-move}
The side move algorithm can be efficiently parallelized. However, this needs to be done carefully to avoid violating detailed balance. Our approach is similar to that used in the \texttt{emcee} package \cite{foreman2013emcee}. Specifically, we divide the ensemble into two groups:
\begin{equation}
    S^{(0)} = \{\mathbf{x}_1, ..., \mathbf{x}_{N/2}\}, \quad S^{(1)} = \{\mathbf{x}_{N/2+1},...,\mathbf{x}_N\}\, .
\end{equation}

At each time step, for each particle in $S^{(0)}$, we randomly select two particles from the complementary set $S^{(1)}$ and perform the side move, applying the Metropolis accept-reject criterion. Then, we follow the same procedure for particles in $S^{(1)}$, selecting particles from the complementary set $S^{(0)}$ to form the side moves and perform Metropolis. The key insight is that we never use particles from the same group as ``sides'' when updating particles within that group, and when we update particles in one group, the particles in another complementary group are treated fixed. This approach preserves the correct detailed balance condition so the parallel version of the sampler keeps $\pi^N$ invariant. We consistently adopt this parallel approach in this paper.

\subsection{Comparison to other affine invariant moves} 
\label{sec-Comparison to other affine invariant moves}
One of the most popular affine invariant sampler is based on \textit{stretch move} \cite{goodman2010ensemble}. In the basic stretch move, at step $m$, one proposes 
\[\tilde{\vx}_i(m) = \vx_j(m) + Z (\vx_i(m) - \vx_j(m))\, ,  \]
where the density of the scaling variable $Z$ satisfies $g(1/z) = zg(z)$, with a commonly used example
\begin{equation}
\label{eqn-density-of-stretch}
    g(z) \propto \begin{cases}
\frac{1}{\sqrt{z}} & \text{if } z \in \left[\frac{1}{a}, a\right] \\
0 & \text{otherwise}
\end{cases}
\end{equation}
where $a$ is the stretch parameter with a recommended default value $a = 2$ (see \cite{foreman2013emcee}). To maintain detailed balance, the proposal is accepted with probability
\[ \mathrm{prob} = \min\{1, Z^{d-1}\frac{\pi(\tilde{\vx}_i(m+1))}{\pi(\mathbf{x}_i(m))}\}\, .\] 

\begin{figure}[ht]
    \centering
    \includegraphics[width=0.9\linewidth]{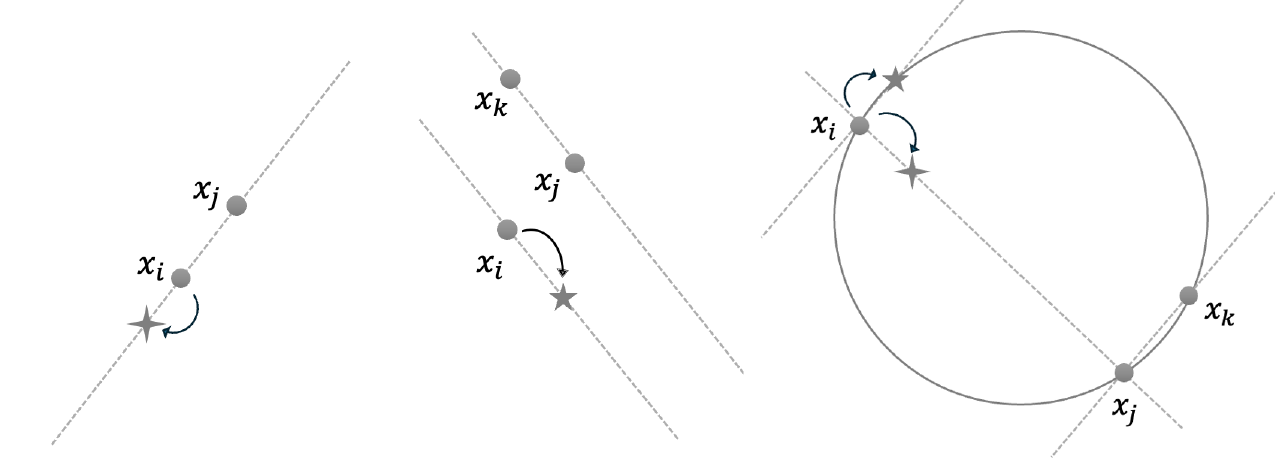}
    \caption{Demonstration of stretch move and side move. Left: stretch move to the four-pointed star; middle: side move to the five-pointed star; right: both moves when points are on a circle}
    \label{fig:stretch-side-moves}
\end{figure}

We illustrate the stretch move and side move concepts in Figure~\ref{fig:stretch-side-moves}. For the stretch move, the new point is positioned along the line formed by $\mathbf{x}_i$ and $\mathbf{x}_j$. In the side move, two distinct particles $\mathbf{x}_j$ and $\mathbf{x}_k$ form a reference side, and $\mathbf{x}_i$ moves in a direction parallel to this side. The effectiveness of each approach depends on the underlying distribution. In the right panel of Figure~\ref{fig:stretch-side-moves}, we illustrate a scenario where points lie on a circle. Here, for the given selection of points, the side move approximately follows the tangential direction, while the stretch move proposes points far from the circle. 

In Section \ref{sec-Formal analysis of high dimensional behaviors}, we provide quantitative high-dimensional scaling analysis for an isotropic Gaussian target. It shows that the side move leads to improved jumped distance compared to the stretch move. The insight is that high-dimensional isotropic Gaussians concentrate on a thin shell, and the inner product between the side direction $\mathbf{x}_j(m) - \mathbf{x}_k(m)$ and $\mathbf{x}_i(m)$ is small with high probability, so the side move proposal typically goes in the favorable tangential direction, in a similar spirit as the right panel of Figure~\ref{fig:stretch-side-moves}.

We note that other affine invariant moves based on empirical covariance exist, such as the \textit{walk move} \cite{goodman2010ensemble}. In the walk move, we select a subset $S$ of particles which are different from $\vx_i(m)$. Denote the empirical mean by $\vm_S$. The proposal is
\begin{equation}
    \tilde{\vx}_i(m+1) = \vx_i(m) + \frac{1}{\sqrt{|S|}}\sum_{j \in S} (\vx_j(m) - \vm_S) \xi_j\, ,
\end{equation}
where $\xi_j \in \cN(0,1)$ are independent normal random variables.
This corresponds to a Gaussian proposal with covariance equal to the empirical covariance of particles in $S$. 

The empirical covariance accounts for global features of the distribution if $|S|$ is large. As discussed in \cite{leimkuhler2018ensemble, reich2021fokker}, local features could be more representative for certain distributions, which may explain the wider popularity of the stretch move in the literature \cite{foreman2013emcee}. We note that when $|S|=2$, the walk move can be shown to be equivalent to the side move with a specific step size. This can be demonstrated by noting:
\begin{equation}
\frac{1}{\sqrt{|S|}}\sum_{j \in S} (\mathbf{x}_j(m) - \mathbf{m}_S) \xi_j = \frac{1}{2\sqrt{2}}(\mathbf{x}_j(m)-\mathbf{x}_k(m))(\xi_j-\xi_k)
\end{equation}
for $S=\{\mathbf{x}_j, \mathbf{x}_k\}$ and $\xi_j - \xi_k \sim \mathcal{N}(0,2)$.

\subsection{Connection to differential evolution} 
\label{sec-Connection to differential evolution}
Affine invariant ensemble MCMC \cite{goodman2010ensemble} was motivated by optimization algorithms. The side move is also connected to methods in the optimization literature, specifically the differential evolution algorithm \cite{storn1997differential}, where particle differences are used to guide the exploration of the function. Differential evolution MCMC \cite{braak2006markov, vrugt2009accelerating} has been developed and has found numerous successful applications.  
In detail, differential evolution MCMC employs the proposal  
\[
\tilde{\vx}_i(m+1) = \vx_i(m) + \gamma (\vx_j(m) - \vx_k(m)) + \sigma \xi\, ,
\]
where $\gamma, \sigma$ are scalars and $\xi \sim \mathcal{N}(0, I_{d\times d})$. The recommended choice for $\gamma$ is $\gamma = \frac{2.38}{\sqrt{2d}}$.  We note that differential evolution MCMC is not generally affine invariant. 

\subsection{Analysis of high dimensional scaling}
\label{sec-Formal analysis of high dimensional behaviors}

    We study the high dimensional scaling of the algorithm for Gaussians at the stationary phase. Since the algorithm is affine invariant, we consider isotropic Gaussians without loss of generality. The analysis suggests $\sigma$ (in side move) and $a-1$ (in stretch move) to scale with $d^{-1/2}$ in high dimensions. We also study the limit of expected squared jumped distance \cite{pasarica2010adaptively,atchade2011towards,roberts2014minimising,yang2020optimal}, which has been used to derive optimal parameters of MCMC. The proof of this proposition can be found in Appendix \ref{Proof-scaling-side-stretch}.
\begin{proposition}
\label{prop-gaussian-acceptance}
    Consider an isotropic Gaussian in $d$ dimensions
    \[\pi(\vx) \propto \exp(-\frac{1}{2}\vx^T\vx)\, , \]
    where $\vx \in \bR^d$. Under the ideal assumption that $\vx_i(m), \vx_j(m), \vx_k(m)$ are all independent draws from this target distribution, the following holds almost surely.
    \begin{itemize}[leftmargin=2em]
        \item For the side move,  $\tilde{\vx}_i(m+1) = \vx_i(m) + \sigma (\vx_j(m)-\vx_k(m)) \xi$, if $\sigma = \frac{\alpha}{\sqrt{d}}$, then we have the following limit of the expected acceptance probability 
        \[\lim_{d \to \infty} \bE[\min\, \{1, \frac{\pi(\tilde{\vx}_i(m+1))}{\pi(\vx_i(m))} \}] =\bE[\min \{1, \exp(-\alpha^2 \xi^2 -  \sqrt{2}\alpha \xi z)\}] \, , \]
        and the expected squared jumped distance
        \[\lim_{d \to \infty} \bE[\|\vx_i(m+1)-\vx_i(m)\|_2^2] =2\alpha^2 \bE[\xi^2 \min \{1, \exp(-\alpha^2 \xi^2 - \sqrt{2}\alpha \xi z \}] \, , \]
        where $\xi \sim \cN(0,1)$ is independent of $z \sim \cN(0,1)$.
        \item For the stretch move, $\tilde{\vx}_i(m+1) = \vx_j(m) + Z (\vx_i(m) - \vx_j(m))$, if $a = 1+\frac{\beta}{\sqrt{d}}$ in \eqref{eqn-density-of-stretch}, then we have the following limit of the expected acceptance probability 
        \[\lim_{d \to \infty} \bE[\min\, \{1, Z^{d-1}\frac{\pi(\tilde{\vx}_i(m+1))}{\pi(\vx_i(m))} \}] =\bE[\min \{1, \exp(-\frac{3}{2}\beta^2U^2 -\sqrt{3}\beta U z)\}] \, , \]
        and the expected squared jumped distance
        \[\lim_{d \to \infty} \bE[\|\vx_i(m+1)-\vx_i(m)\|_2^2] =2\beta^2 \bE[U^2\min \{1, \exp(-\frac{3}{2}\beta^2U^2 -\sqrt{3}\beta U z)\}] \, , \]
        where $U \sim \mathrm{Unif}[-1,1]$ is independent of $z \sim \cN(0,1)$.
    \end{itemize}
\end{proposition}
The scaling of $\sigma \sim d^{-1/2}$ for the side move is natural since it resembles a basic random walk, aligning with existing scaling results for non-affine-invariant random walk Metropolis \cite{gelman1997weak, yang2020optimal}. The above result applies to arbitrary Gaussians given the affine invariance property of the side move. The scaling $a - 1 \sim d^{-1/2}$ for the stretch move is perhaps less obvious. The presence of the $Z^{d-1}$ factor is the key element leading to such scaling.

The proposition demonstrates that affine invariant ensemble samplers lead to the same high-dimensional scaling as single-chain random walk (at least for Gaussian targets). However, the ensemble methods maintain consistent performance for highly anisotropic distributions, while basic random walk without affine invariance does not.

\begin{figure}[ht]
    \centering
    \includegraphics[width=0.95\linewidth]{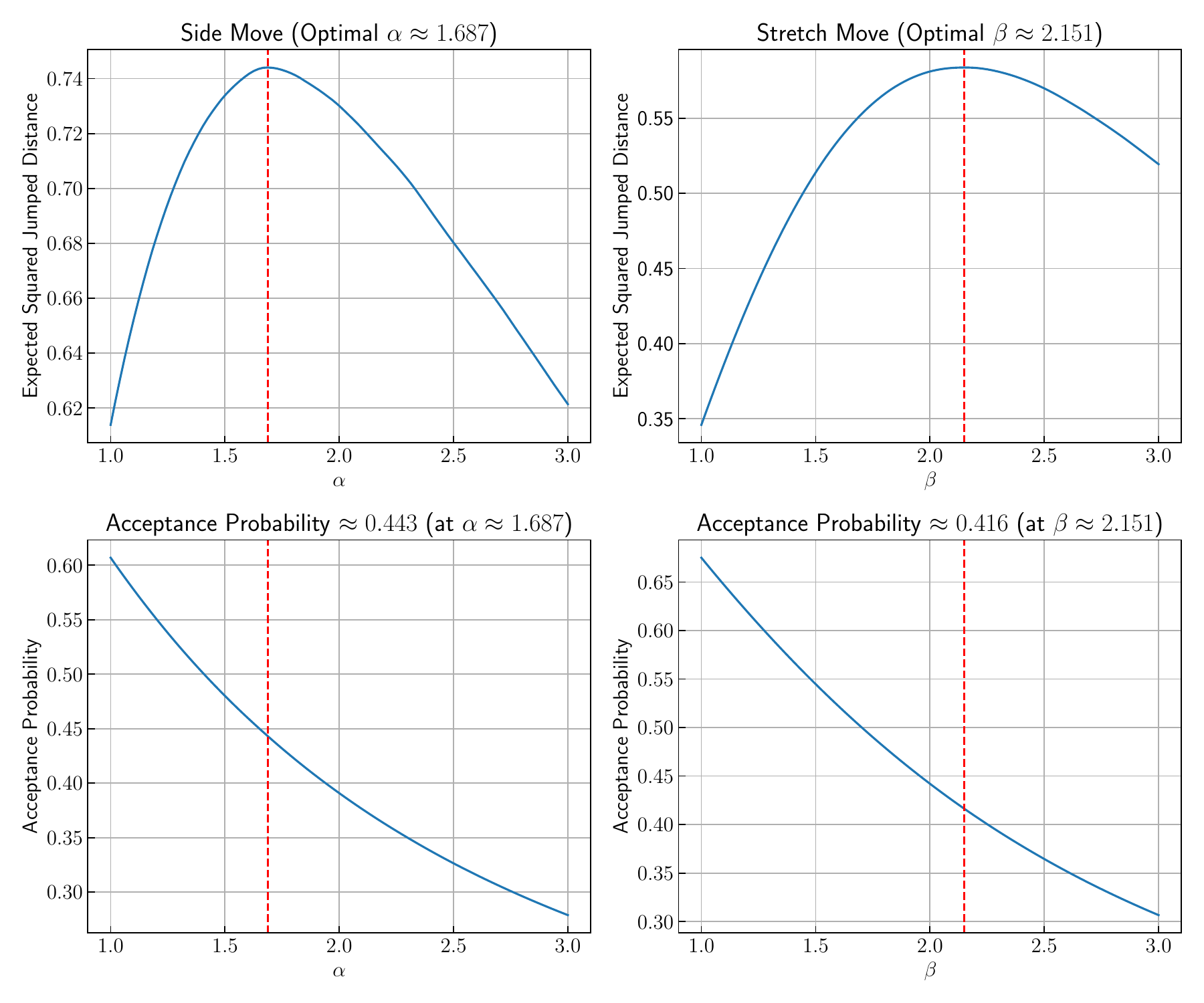}
    \caption{Side move versus stretch move: expected acceptance rate and squared expected jumped distance. Optimal $\alpha$ and $\beta$ in terms of the squared expected jumped distance are marked.}
    \label{fig: max expected length}
\end{figure}

With Proposition \ref{prop-gaussian-acceptance}, we can perform simulations to find the optimal $\alpha$ and $\beta$ that lead to the largest expected squared jumped distance. The results are shown in Figure \ref{fig: max expected length}. In three decimal numbers, we found that the optimal $\alpha$ is approximately $1.687$ with a squared jumped distance of $0.744$, and the optimal $\beta$ is approximately $2.151$ with a squared jumped distance of $0.584$. The corresponding acceptance probabilities are $0.443$ and $0.416$, respectively. With these parameters, the side move can achieve approximately $0.744/0.584 \approx 1.27$ times the squared jumped distance compared to the stretch move. 

The random walk scaling shows that the sampler requires $O(d)$ steps to traverse the distribution's support. We conduct numerical experiments in Section \ref{sec-exp-gaussian} for Gaussian targets using the optimal parameters described above. These experiments demonstrate that the autocorrelation time scales as $O(d)$. The side move achieves a smaller constant in the $O(d)$ scaling of autocorrelation time compared to the stretch move.
\begin{newremark}
    We note that using a different distribution for $\xi$ in the side move results in different optimal parameters. For example, if we choose $\xi \sim \mathrm{Unif}[-1,1]$, by examining the formula, we find that the optimal $\alpha$ and $\beta$ are related by $\alpha = \sqrt{\frac{3}{2}}\beta$. With this choice, the side move achieves exactly $1.5$ times the expected squared jumped distance compared to the stretch move.
\end{newremark}

\section{Derivative-based Affine Invariant HMC Samplers}
\label{sec-Affine Invariant Hamiltonian Monte Carlo Samplers}
The samplers in the previous section are derivative-free, making them simple, convenient, and widely used in applications. However, when derivative information is available, derivative-based samplers such as Langevin \cite{roberts1996exponential} and Hamiltonian Monte Carlo (HMC) \cite{neal2011mcmc} have been shown to scale better with dimension \cite{roberts1998optimal,beskos2013optimal}.

There has been considerable interest in developing affine invariant Langevin samplers based on Riemannian geometry and empirical covariance preconditioning \cite{girolami2011riemann,greengard2015ensemblized,leimkuhler2018ensemble, garbuno2020interacting,garbuno2020affine}. To the best of our knowledge, it remains unknown whether an adjusted ensemble HMC that is affine invariant can be efficiently developed. Since HMC is one of the state-of-the-art samplers that overcomes random walk-like behavior and scales favorably with dimension, developing an affine invariant ensemble HMC is important for further understanding and advancement of affine invariant samplers. This section aims to discuss the design principles and propose new samplers of this kind.

\subsection{Hamiltonian Monte Carlo} Let $\pi \propto \exp(-V)$ where $V: \bR^d \to \bR$ is a potential function. In HMC, we consider a probability density $\exp(-V(\vx)-\frac{1}{2}\vp^TM^{-1}\vp)$ over an extended state space $(\vx, \vp)$ where $\vx, \vp \in \bR^d$ and $\vp$ is often referred to as the momentum vector; $M$ is the called the mass matrix which often plays the role of preconditioning.

The following Hamiltonian dynamics keep the joint distribution invariant
\begin{equation*}
    \begin{aligned}
        \frac{{\rm d}\vx}{{\rm d}t} =  M^{-1}\vp, \quad \frac{{\rm d}\vp}{{\rm d}t} = -\nabla V(\vx)\, .\\
    \end{aligned}
\end{equation*}
The standard HMC sampler alternates between two operations: Gibbs sampling to refresh the momentum $\vp \sim \cN(0,M)$ and running approximate Hamiltonian dynamics followed by momentum negation with Metropolis correction. Since both operations individually preserve the invariant distribution, their composition does as well.

More specifically, let $\sfL_h$ be the deterministic map that performs one leapfrog step for the Hamiltonian dynamics with time step size $h$. That is, $(\vx_h, \vp_h) = \sfL_h (\vx, \vp)$ satisfies
\begin{equation}
    \vp_{h/2} = \vp - \frac{h}{2}\nabla V(\vx), \quad \vx_h = \vx + hM^{-1}\vp_{h/2}, \quad \vp_h = \vp_{h/2} - \frac{h}{2}\nabla V(\vx_h)\, .
\end{equation}
We denote $\sfP$ as the momentum flip operator such that $\sfP(\vx,\vp) = (\vx,-\vp)$. Then, each single iteration of HMC goes as follows:
\begin{itemize}[leftmargin=2em]
    \item Sample a momentum $\vp \sim \cN(0,M)$.
    \item Propose an update $(\tilde{\vx}, \tilde{\vp}) = \sfP \sfL_h^n (\vx, \vp)$ where $n$ is the number of leapfrog steps taken; such proposal is accepted with probability
\[ \text{prob} = \min\,\{1,\exp(-V(\tilde{\vx})-\frac{1}{2}\tilde{\vp}^TM^{-1}\tilde{\vp} + V(\vx) + \frac{1}{2}\vp^T M^{-1}\vp)\}\, . \]
\end{itemize}
The key to the derivation of HMC is the property that $\sfP \sfL_h^n = (\sfP \sfL_h^n)^{-1}$ so the detailed balance for the second step is valid. Modern development of HMC focuses on tuning the stepsize and integration time \cite{hoffman2014no,carpenter2017stan}, among many others. 

\subsection{Covariance preconditioning and its challenge}
How shall we develop an ensemble HMC sampler that is affine invariant? The simplest idea is to use empirical covariance as the inverse mass matrix $M$ for preconditioning. To implement this idea, let us follow the ensemble splitting strategy in Section \ref{sec-parallel-side-move} that is easy to parallelize and works better with detailed balance. Consider an ensemble of $N$ particles $\vx_1, ..., \vx_N$ split into two groups  
\begin{equation}
\label{eqn-split-ensembles}
    S^{(0)} = \{\mathbf{x}_1, ..., \mathbf{x}_{N/2}\}, \quad S^{(1)} = \{\mathbf{x}_{N/2+1},...,\mathbf{x}_N\}\, ,
\end{equation}
and the joint distribution
\[ \exp\left(-\sum_{i=1}^{N/2} \left(V(\vx_i) + \frac{1}{2}\vp_i^{T}\mathrm{Cov}_{S^{(1)}}\vp_i\right) -  \sum_{i=N/2+1}^{N} \left(V(\vx_i) + \frac{1}{2}\vp_i^{T}\mathrm{Cov}_{S^{(0)}}\vp_i\right)\right)\, , \]
where $\vp_i \in \bR^d$ is the associated momentum vector with $\vx_i$. Here 
$\mathrm{Cov}_{S^{(0)}}$ and $\mathrm{Cov}_{S^{(1)}}$ are the empirical covariance matrices of particles in $S^{(0)}$ and $S^{(1)}$. The formula is
\[\mathrm{Cov}_{S^{(0)}} = \frac{2}{N}\sum_{i=1}^{N/2} (\vx_i - \vm_{S^{(0)}})(\vx_i - \vm_{S^{(0)}})^T\]
where $\vm_{S^{(0)}}$ is the empirical mean for particles in $S^{(0)}$.

We note that when marginalized over all $\vp_i$, the above joint distribution does not give the correct marginal distribution for $\vx_i$ due to the existence of normalization constants. In fact, we need to additionally add $-\frac{N}{4}\log\det \mathrm{Cov}_{S^{(0)}} - \frac{N}{4}\log\det \mathrm{Cov}_{S^{(1)}}$ to the potential to address this issue. The modified final joint distribution is
\begin{equation}
    \pi^N_\star(\vx_1,...,\vx_N, \vp_1,...\vp_N) \propto \exp(-V_\star(\vx_1,...,\vx_N, \vp_1,...\vp_N))
\end{equation} where 
\begin{equation}
    \begin{aligned}
        V_\star(\vx_1,...,\vx_N, \vp_1,...\vp_N) = & \sum_{i=1}^{N/2} \left(V(\vx_i) + \frac{1}{2}\vp_i^{T}\mathrm{Cov}_{S^{(1)}}\vp_i\right) \\
         & +  \sum_{i=N/2+1}^{N} \left(V(\vx_i) + \frac{1}{2}\vp_i^{T}\mathrm{Cov}_{S^{(0)}}\vp_i\right) \\
         & - \frac{N}{4}\log\det \mathrm{Cov}_{S^{(0)}} - \frac{N}{4}\log\det \mathrm{Cov}_{S^{(1)}}\, .
    \end{aligned}
\end{equation}
A direct calculation shows that the gradient of the added potential is  \[\nabla_{\vx_i} \log \det \mathrm{Cov}_{S^{(0)}} =  \mathrm{Cov}^{-1}_{S^{(0)}} \nabla_{
\vx_i}\mathrm{Cov}_{S^{(0)}} =   \frac{4}{N}\mathrm{Cov}^{-1}_{S^{(0)}}(\vx_i - \vm_{S^{(0)}})\, .\]
 Thus, using the HMC pipeline, we obtain the following algorithm:
\begin{itemize}[leftmargin=2em]
    \item We fix particles in $S^{(1)}$, and update particles in $S^{(0)}$: for each $1\leq i \leq N/2$, we sample $\vp_i \sim \cN(0, \mathrm{Cov}_{S^{(1)}}^{-1})$, run leapfrog approximation of the dynamics
    \begin{equation*}
    \begin{aligned}
        \frac{{\rm d}\vx_i}{{\rm d}t} =  \mathrm{Cov}_{S^{(1)}}\vp_i, \quad \frac{{\rm d}\vp_i}{{\rm d}t} = -\nabla V(\vx_i) + \mathrm{Cov}^{-1}_{S^{(0)}}(\vx_i - \vm_{S^{(0)}})\, ,
    \end{aligned}
\end{equation*}
and apply the Metropolis accept-reject criterion. 


\item Then we fix particles in $S^{(0)}$, and update particles in $S^{(1)}$ in a similar fashion.
\item Iterate the above two steps.
\end{itemize}
We can show the algorithm is affine invariant, thanks to the covariance preconditioning. In fact, given an invertible affine transformation $\phi(\mathbf{x}) = A\mathbf{x} + \mathbf{b}$, we can transform
\begin{equation}
\begin{aligned}
    &(\vx_1,...,\vx_N) \to (\vy_1,...,\vy_N) = (A\vx_1+\mathbf{b},...,A\vx_N+\mathbf{b})\\
    & (\vp_1,...,\vp_N) \to (\vr_1,...,\vr_N) = (A^{-T}\vp_1,...,A^{-T}\vp_N)\, .
\end{aligned}
\end{equation}
By the change of variables, we get (for $1\leq i \leq N/2$)
 \begin{equation*}
    \begin{aligned}
        \frac{{\rm d}\vy_i}{{\rm d}t} =  \mathrm{Cov}_{S^{(1)}_\vy}\vr_i, \quad \frac{{\rm d}\vr_i}{{\rm d}t} = -\nabla V^\phi(\vy_i) + \mathrm{Cov}^{-1}_{S^{(0)}_\vy}(\vy_i - \vm_{S^{(0)}_\vy})\, ,
    \end{aligned}
\end{equation*}
where $V^\phi(\vy) = V(\phi^{-1}(\vy))$ corresponds to the potential of the transformed density $\phi \# \pi$, and $S^{(0)}_\vy$ is the first group of transformed particles. One can further show that the above correspondence holds for the leapfrog scheme, and the final acceptance ratio remains the same for the original and transformed dynamics. This demonstrates the affine invariance property.

However, we note that here the $N/2$ particles in a group will need to be accepted or rejected \textit{at the same time}, as they are coupled through the $\log\det$ term. The potential simultaneous rejection can be wasteful and inefficient.  A different approach is therefore needed to make an efficient ensemble HMC with affine invariance.

\subsection{Hamiltonian walk move sampler} 
Our discussion in the previous subsection suggests it is preferable to decouple the momentum for different particles in the joint distribution. Now consider the simple joint distribution with identity mass matrices:
\[ \exp\left(-\sum_{i=1}^{N/2} \left(V(\vx_i) + \frac{1}{2}\vp_i^{T}\vp_i\right) -  \sum_{i=N/2+1}^{N} \left(V(\vx_i) + \frac{1}{2}\vp_i^{T}\vp_i\right)\right)\, , \]
where again, the particles are split into two groups \begin{equation}
    S^{(0)} = \{\mathbf{x}_1, ..., \mathbf{x}_{N/2}\}, \quad S^{(1)} = \{\mathbf{x}_{N/2+1},...,\mathbf{x}_N\}\, .
\end{equation}
Instead of running the standard Hamiltonian dynamics, we note that with a preconditioning matrix $B \in \bR^{d\times d}$, the preconditioned dynamics
\begin{equation*}
    \begin{aligned}
        \frac{{\rm d}\vx_i}{{\rm d}t} =  B\vp_i, \quad \frac{{\rm d}\vp_i}{{\rm d}t} = -B^T\nabla V(\vx_i)\, 
    \end{aligned}
\end{equation*}
will also preserve the joint distribution. A similar antisymmetric preconditioning has been explored in \cite{leimkuhler2018ensemble} for underdamped Langevin dynamics. 

Our goal is to select an appropriate $B$ that depends on the ensembles $S^{(0)}$ and $ S^{(1)}$ to make the preconditioned Hamiltonian dynamics affine invariant. We must choose $B$ in a way that enables efficient calculation of the acceptance rate. Another important observation is that the dimension of $\mathbf{p}_i$ need not match the dimension of $\mathbf{x}_i$; in such cases, the matrix $B$ is chosen to be rectangular to ensure dimensional consistency.

Based on the above insights, we propose the following algorithm.
\begin{itemize}[leftmargin=2em]
    \item We fix particles in $S^{(1)}$, and update particles in $S^{(0)}$: for each $1\leq i \leq N/2$, we sample $\vp_i \sim \cN(0, I_{N/2\times N/2})$, and run $n$ steps of leapfrog approximation of the dynamics
    \begin{equation*}
    \begin{aligned}
        \frac{{\rm d}\vx_i}{{\rm d}t} = B_{S^{(1)}} \vp_i, \quad \frac{{\rm d}\vp_i}{{\rm d}t} = -B_{S^{(1)}}^T\nabla V(\vx_i)\, ,
    \end{aligned}
\end{equation*}
where 
\begin{equation}
\label{eqn-centered-ensemble-S1}
    B_{S^{(1)}} = \frac{1}{\sqrt{N/2}}[\vx_{N/2+1} - \vm_{S^{(1)}},...,\vx_N - \vm_{S^{(1)}}] \in \bR^{d \times N/2}
\end{equation}
is called a normalized centered ensemble for particles in group $S^{(1)}$. Here $\vm_{S^{(1)}}$ is the mean of all particles in $S^{(1)}$.

More precisely, denote $\sfL_{h, B_{S^{(1)}}}$ as the corresponding leapfrog operator. That is, $(\vx_h, \vp_h) = \sfL_{h, B_{S^{(1)}}} (\vx, \vp)$ satisfies
\begin{equation}
\label{eqn-leapfrog-H-walk-move}
    \vp_{h/2} = \vp - \frac{h}{2}B_{S^{(1)}}^T\nabla V(\vx), \quad \vx_h = \vx + hB_{S^{(1)}}\vp_{h/2}, \quad \vp_h = \vp_{h/2} - \frac{h}{2}B_{S^{(1)}}^T\nabla V(\vx_h)\, .
\end{equation}
We propose $(\tilde{\vx}_i, \tilde{\vp}_i) = \sfP\sfL_{h, B_{S^{(1)}}}^n (\vx_i, \vp_i)$, and we accept this proposal for each $1\leq i \leq N/2$ with probability 
\[\text{prob}_i = \min\, \{1, \exp(-V(\tilde{\vx}_i)-\frac{1}{2}\tilde{\vp}_i^T\tilde{\vp}_i + V(\vx_i) + \frac{1}{2}\vp_i^T\vp_i)\}\, . \]
\item Then, we fix particles in $S^{(0)}$, and update particles in $S^{(1)}$ in a similar fashion. Here the dynamics for particles $N/2+1 \leq i \leq N$ are
\begin{equation*}
    \begin{aligned}
        \frac{{\rm d}\vx_i}{{\rm d}t} = B_{S^{(0)}} \vp_i , \quad \frac{{\rm d}\vp_i}{{\rm d}t} = -B_{S^{(0)}}^T\nabla V(\vx_i)\, ,
    \end{aligned}
\end{equation*}
with 
\begin{equation}
\label{eqn-centered-ensemble-S0}
    B_{S^{(0)}} = \frac{1}{\sqrt{N/2}}[\vx_{1} - \vm_{S^{(0)}},...,\vx_{N/2} - \vm_{S^{(0)}}] \in \bR^{d \times N/2}\, .
\end{equation}
\item Iterate the above two steps.
\end{itemize}

The algorithm is affine invariant. Indeed, given an invertible affine transformation $\phi(\mathbf{x}) = A\mathbf{x} + \mathbf{b}$, we can transform
\begin{equation}
\begin{aligned}
    (\vx_1,...,\vx_N) \overset{\phi}{\to} (\vy_1,...,\vy_N) = (A\vx_1+\mathbf{b},...,A\vx_N+\mathbf{b})
\end{aligned}
\end{equation}
while keeping $\vp_i$ untransformed. This differs from the previous subsection where both $\vx$ and $\vp$ are transformed. However, this difference does not matter because our ultimate goal is to sample the distribution on 
 $\vx$, not $\vp$. At the continuous level, we obtain
\begin{equation*}
    \begin{aligned}
        \frac{{\rm d}\vy_i}{{\rm d}t} = B_{S^{(1)}_\vy} \vp_i, \quad \frac{{\rm d}\vp_i}{{\rm d}t} = -B_{S^{(1)}_\vy}^T\nabla V^\phi(\vy_i)\, ,
    \end{aligned}
\end{equation*}
which represents the preconditioned Hamiltonian dynamics applied to the transformed density $\phi \# \pi$. This relationship extends to the leapfrog discretization of the dynamics. Furthermore, the acceptance ratio remains identical for both the transformed and untransformed cases, thus confirming the method's affine invariance.

We note that the centered ensembles here play a role similar to the Cholesky factor of the empirical covariance matrix. While it is possible to directly use the Cholesky factor for preconditioning, which would result in a momentum vector with the same dimension as the state, using centered ensembles avoids the computational cost of factorization. Similar use of centered ensembles to avoid Cholesky has been considered in \cite{garbuno2020affine} for covariance preconditioned Langevin dynamics.

The algorithm may be interpreted as a gradient-enhanced walk move. In one form of the walk move \cite{goodman2010ensemble}, the proposal is $\mathbf{x}_i +  B_{S^{(1)}} \mathbf{z}$ with $\mathbf{z} \sim \mathcal{N}(0,I_{N/2\times N/2})$. In our algorithm, the derivative of $\mathbf{x}_i$ is similarly a linear combination of the columns of $B_{S^{(1)}}$. When the integration time is small, the leading order term follows the same direction as $B_{S^{(1)}} \mathbf{z}$. With longer integration times, our approach adaptively determines the move strength using Hamiltonian-type dynamics that incorporate derivative information of the potential. For this reason, we term the method the \textit{Hamiltonian walk move} sampler. In general, similar to the walk move sampler, we can choose an arbitrary subset $S$ from the complementary ensemble to form the centered ensemble; in the above algorithm, $S$ is chosen to contain all particles in the complementary ensemble. Choosing a smaller $S$ could reduce the arithmetic cost.

\subsection{Hamiltonian side move sampler}
In the last subsection, the use of empirical covariance or centered ensembles of all particles in the complementary ensemble makes the algorithm affine invariant. This approach accounts for global statistics through covariance. 
Alternatively, we can use the local side move direction to derive an affine invariant algorithm as follows.
\begin{itemize}[leftmargin=2em]
    \item We fix particles in $S^{(1)}$, and update particles in $S^{(0)}$: for each $1\leq i \leq N/2$, we sample $p_i \sim \cN(0, 1)$ and two particles $\vx_j$ and $\vx_k$ from $S^{(1)}$. We run $n$ steps of leapfrog approximation of the dynamics
    \begin{equation*}
    \begin{aligned}
        \frac{{\rm d}\vx_i}{{\rm d}t} = \frac{1}{\sqrt{2d}}(\vx_j-\vx_k) p_i, \quad \frac{{\rm d}p_i}{{\rm d}t} = -\frac{1}{\sqrt{2d}}(\vx_j-\vx_k) ^T\nabla V(\vx_i)\, .
    \end{aligned}
\end{equation*}

More precisely, for the $i$-th particle, denote $\sfL_{h, (\vx_j-\vx_k)/\sqrt{2d}}$ as the corresponding leapfrog operator. That is, $(\vx_h, p_h) = \sfL_{h, (\vx_j-\vx_k)/\sqrt{2d}} (\vx, p)$ satisfies
\begin{equation}
\label{eqn-H-side-move-leapfrog}
    \begin{aligned}
p_{h/2} &= p - \frac{h}{2}\left(\frac{\mathbf{x}_j-\mathbf{x}_k}{\sqrt{2d}}\right)^T\nabla V(\mathbf{x}), \\
\mathbf{x}_h &= \mathbf{x} + h\left(\frac{\mathbf{x}_j-\mathbf{x}_k}{\sqrt{2d}}\right)p_{h/2}, \\
p_h &= p_{h/2} - \frac{h}{2}\left(\frac{\mathbf{x}_j-\mathbf{x}_k}{\sqrt{2d}}\right)^T\nabla V(\mathbf{x}_h)\, .
\end{aligned}
\end{equation}

We propose $(\tilde{\vx}_i, \tilde{p}_i) = \sfP\sfL_{h, (\vx_j-\vx_k)/\sqrt{2d}}^n (\vx_i, p_i)$, and we accept the proposal with probability 
\[\text{prob}_i = \min\, \{1, \exp(-V(\tilde{\vx}_i)-\frac{1}{2}\tilde{p}_i^2 + V(\vx_i) + \frac{1}{2}p_i^2)\}\, . \]
\item Then, we fix particles in $S^{(0)}$, and update particles in $S^{(1)}$ in a similar fashion. Here for each $N/2+1 \leq i \leq N$, the corresponding $\vx_j$ and $\vx_k$ are randomly drawn from $S^{(0)}$.
\item Iterate the above two steps.
\end{itemize}

This approach only needs to calculate directional gradients rather than full gradients. We refer to the algorithm as the \textit{Hamiltonian side move} sampler since it uses precisely the same direction as the standard side move, with the move length adaptively determined by Hamiltonian-type dynamics. Like our previous methods, this approach maintains affine invariance for the same underlying reasons. Similar to discussions at the end of Section \ref{sec-Comparison to other affine invariant moves}, the Hamiltonian side move may be interpreted as a specific local variant (i.e., $|S|=2$) of the Hamiltonian walk move.

\subsection{Analysis of high dimensional scaling}
\label{sec-Analysis of high dimensional scaling}
We study the stepsize scaling in high dimensions for our proposed affine invariant HMC algorithms. As in Section \ref{sec-Formal analysis of high dimensional behaviors}, we focus on isotropic Gaussian distributions and the stationary phase of the algorithms. We denote by $T$ a fixed time horizon for the Hamiltonian dynamics. All statements below pertain to a single particle $\mathbf{x}_i$ during one iteration of the algorithm. Following our convention from previous sections, we denote $\mathbf{x}_i(m)$ as the state at step $m$, and $\mathbf{x}_i(m+1)$ as the state in the subsequent iteration after applying the preconditioned Hamiltonian dynamics and the accept-reject mechanism. For notational simplicity, we omit the indices $m$ and $m+1$ when the context is clear.
The proof of the proposition can be found in Appendix \ref{Proof-scaling-HMC}.
\begin{proposition}
\label{thm: scaling-HMC}
    Consider an isotropic Gaussian in $d$ dimensions
    \[\pi(\vx) \propto \exp(-\frac{1}{2}\vx^T\vx)\, , \]
    where $\vx \in \bR^d$. Under the ideal assumption that all $\vx_i$ are independent draws from this target distribution, the following holds almost surely.
    \begin{itemize}[leftmargin=2em]
        \item For the Hamiltonian walk move, if we take $h =  \alpha d^{-1/4}, n = T/h$ and assume $\lim_{d\to \infty} \frac{d}{N(d)/2} = \rho \in [0,1)$, then as $d\to\infty$, the acceptance probability converges to 
\[ \bE[\min\{1, \exp(\cN(\alpha^4 \mu_{\rho}, \alpha^4\sigma_{\rho}))\}] \, , \]
where $\cN(\alpha^4 \mu_{\rho}, \alpha^4\sigma_{\rho})$ is a Gaussian distribution with 
\[\mu_{\rho} = -\frac{1}{32}\int \lambda^{4} \sin^2(\sqrt{\lambda} T) {\rm d}\nu_{\rho}(\lambda),\quad \sigma_\rho = \frac{1}{16}\int \lambda^{6} \sin^2(\sqrt{\lambda} T) {\rm d}\nu_{\rho}(\lambda)\, .\]
For $\rho \in [0,1)$, 
$\mathrm{d}\nu_{\rho}(\lambda) = \frac{1}{2\pi \rho \lambda} \sqrt{(c-\lambda)(\lambda-b)}\chi_{[b,c]}(\lambda)\mathrm{d}\lambda$, 
where $b = (1-\sqrt{\rho})^2, c = (1+\sqrt{\rho})^2$ and $\chi_{[b,c]}$ is the characteristic function of $[b,c]$. When $\rho = 0$, $\mathrm{d}\nu_{\rho}(\lambda)$ concentrates on the Dirac mass at $\lambda = 1$. Moreover, the expected squared jumped distance satisfies
\[\lim_{d \to \infty} \frac{1}{d}\bE[\|\vx_i(m+1)-\vx_i(m)\|_2^2] = 4\int \sin^2(\frac{\sqrt{\lambda} T}{2}){\rm d}\nu_{\rho}(\lambda)\bE[\min\{1, \exp(\cN(\alpha^4 \mu_{\rho}, \alpha^4\sigma_{\rho}))\}]\, . \]
        \item For the Hamiltonian side move, we can take $h =  \alpha < 2$ to be a constant that does not depend on $d$. Let $n$ be the number of leapfrog steps. As $d\to \infty$, the acceptance probability converges to a nonzero limit $\bE[\min\{1, \exp(P(z_1,z_2,n,\alpha))\}]$ where 
\[P(z_1,z_2,n,\alpha) = \frac{\alpha^2}{8}\left(\sin^2(n\phi)(z_1^2 - z_2^2) - \sin(2n\phi)z_1z_2\right)\, .\]
Here $z_1 \sim \cN(0,1), z_2\sim \cN(0,\frac{1}{1-\alpha^2/4})$ are independent, and $\phi \in [0,\pi]$ satisfies $\cos \phi = 1 - \frac{\alpha^2}{2}$.
Moreover, the expected squared jumped distance satisfies
\[\lim_{d \to \infty} \bE[\|\vx_i(m+1)-\vx_i(m)\|_2^2] = \bE[Q(z_1,z_2, n, \alpha)\min\{1, \exp(P(z_1,z_2,n,\alpha))\} ]\, , \]
where we define
\[Q(z_1,z_2, n, \alpha) = (\cos(n\phi)-1)^2z_1^2 + \sin^2(n\phi)z_2^2 + 2(\cos(n\phi)-1)\sin(n\phi)z_1z_2\, .\]
    \end{itemize}
\end{proposition}
The proposition demonstrates that in $d$ dimensions, the expected squared distance traveled in one iteration of the Hamiltonian walk move is $O(d)$, whereas it is only $O(1)$ for the Hamiltonian side move since the latter restricts movement along a single line in each iteration.

Overall, the Hamiltonian walk move requires $O(d^{1/4})$ leapfrog steps, or gradient and function evaluations, to traverse the support of the target distribution. This is much more efficient than the $O(d)$ evaluations needed for the Hamiltonian side move and the previously developed derivative-free stretch and side moves.

\section{Numerical Experiments}
\label{sec-Numerical Experiments}


\subsection{Evaluation criterion} 
We investigate the efficiency of affine invariant ensemble samplers through numerical experiments. Our main evaluation criterion is the autocorrelation time at the stationary phase, following \cite{goodman2010ensemble}. These samplers generate sequences $(\mathbf{x}_1(m),...,\mathbf{x}_N(m))$ for $1 \leq m \leq M$, where $M$ represents the length of the ensemble chain. We use these ensembles to estimate the observable
\[A = \mathbb{E}^{\mathbf{x} \sim \pi}[f(\mathbf{x})] = \int f(\mathbf{x}) \pi(\mathbf{x}){\rm d}\mathbf{x}\, ,  \]
via the approximation
\[\hat{A}_e = \frac{1}{M}\sum_{m=1}^{M} F(\mathbf{x}_1(m),...,\mathbf{x}_N(m)) =\frac{1}{M}\sum_{m=1}^{M}  \left(\frac{1}{N}\sum_{i=1}^N f(\mathbf{x}_i(m))\right)\, .  \]
At the stationary phase, for large $M$, the variance of the estimator satisfies
\[\mathrm{Var}(\hat{A}_e) \approx \frac{\tau_e}{M}\mathrm{Var}^{\mathbf{x}_1,...,\mathbf{x}_N \sim \pi^N}[F(\mathbf{x}_1,...,\mathbf{x}_N)] = \frac{\tau_e}{NM}\mathrm{Var}^{\mathbf{x} \sim \pi}[f(\mathbf{x})]\, , \]
where $\tau_e$ is the integrated autocorrelation time for the ensemble method defined as
$\tau_e = \sum_{m=-\infty}^{+\infty} \frac{C_e(m)}{C_e(0)}$
with the autocovariance function defined as \[C_e(m) = \lim_{m' \to \infty} \mathrm{Cov}[F(\vx_1(m'),...,\vx_N(m')), F(\vx_1(m+m'),...,\vx_N(m+m'))]\, .\]
The autocorrelation function at lag $m$ is the ratio $\frac{C_e(m)}{C_e(0)}$.

In contrast, when using a single chain MCMC algorithm such as HMC, we obtain a single sequence $\mathbf{x}(m)$ for $1\leq m \leq M$. The estimator becomes
\[\hat{A}_s = \frac{1}{M}\sum_{m=1}^{M} f(\mathbf{x}(m))\, ,  \]
with variance \[\mathrm{Var}(\hat{A}_s) \approx \frac{\tau_s}{M}\mathrm{Var}^{\mathbf{x} \sim \pi}[f(\mathbf{x})]\, ,\] where $\tau_s = \sum_{m=-\infty}^{+\infty} \frac{C_s(m)}{C_s(0)}$ 
and $C_s(m) = \lim_{m' \to \infty} \mathrm{Cov}[f(\vx(m')), f(\vx(m+m'))]$. As noted by \cite{goodman2010ensemble}, a natural criterion to compare performance is the values of $\tau_e$ and $\tau_s$. The estimation procedure of the autocorrelation time follows \cite[Section 5]{goodman2010ensemble}.


Code is available at \url{https://github.com/yifanc96/AffineInvariantSamplers}.
\subsection{Synthetic: Gaussian}
\label{sec-exp-gaussian}
As our first example, we consider Gaussian distributions. Given a dimension $d$ and condition number $\kappa$, we generate $d$ eigenvalues equi-distributed between $10^{-1}$ and $10^{-1}\kappa$. These form a diagonal matrix $\Sigma$ with diagonal entries equal to the eigenvalues, which we use as the precision matrix of the Gaussian distribution. 


For all ensemble samplers, we use $N=2d$ walkers. For the stretch move, we use parameter $a = 1 + 2.151d^{-1/2}$ as suggested in Section \ref{sec-Formal analysis of high dimensional behaviors}. For the side move, we set $\sigma = 1.687d^{-1/2}$. For HMC, we fix the total time horizon at $T=1$ and use $n=2$ or $n=10$ leapfrog steps with corresponding step size $h=T/n$. The same parameters apply to the Hamiltonian walk and side moves.

We run $2\times 10^5$ iterations of these samplers as burn-in, followed by $10^6$ steps treated as the stationary phase. The autocorrelation function\footnote{Due to memory constraints, we often thin the samples by a factor of $10$, compute the autocorrelation function and time for these thinned samples. We explicitly note this whenever the thinning is used.} is computed for the observable $f(\vx)=x_1$, the first component.

\begin{figure}
    \centering
    \includegraphics[width=0.48\linewidth]{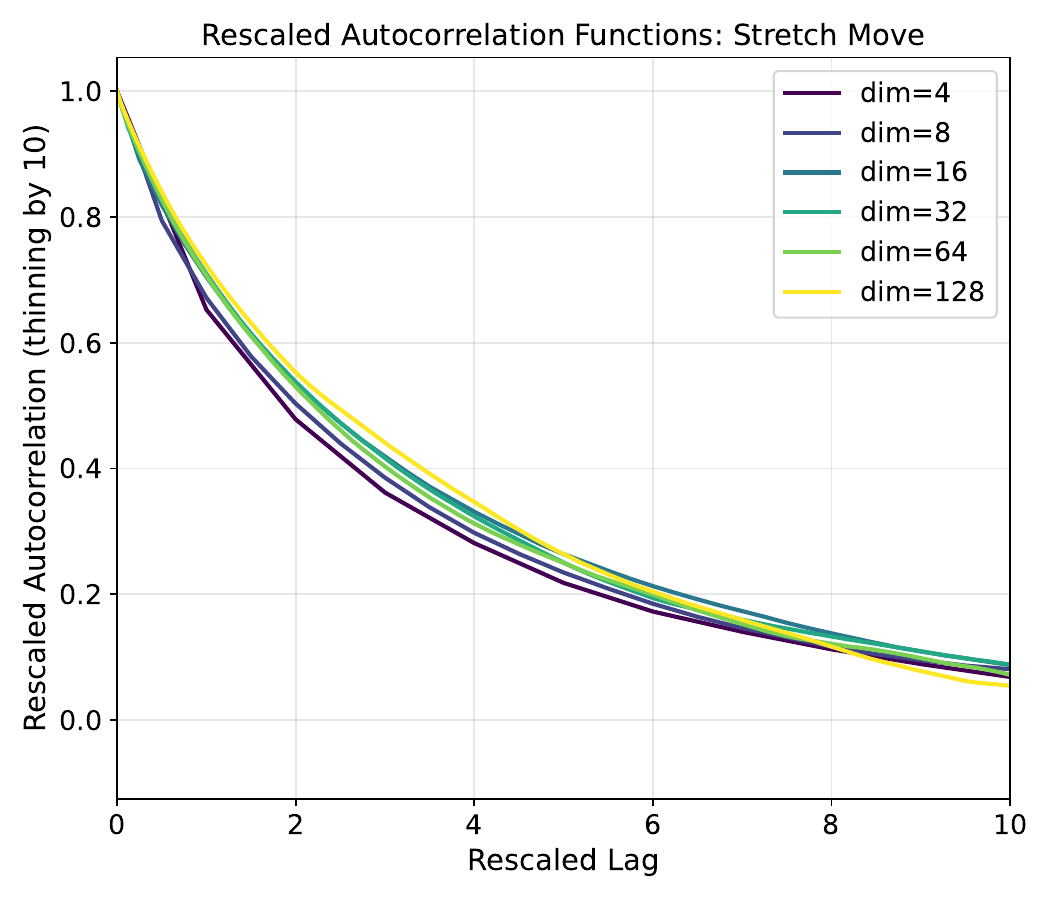}
    \includegraphics[width=0.48\linewidth]{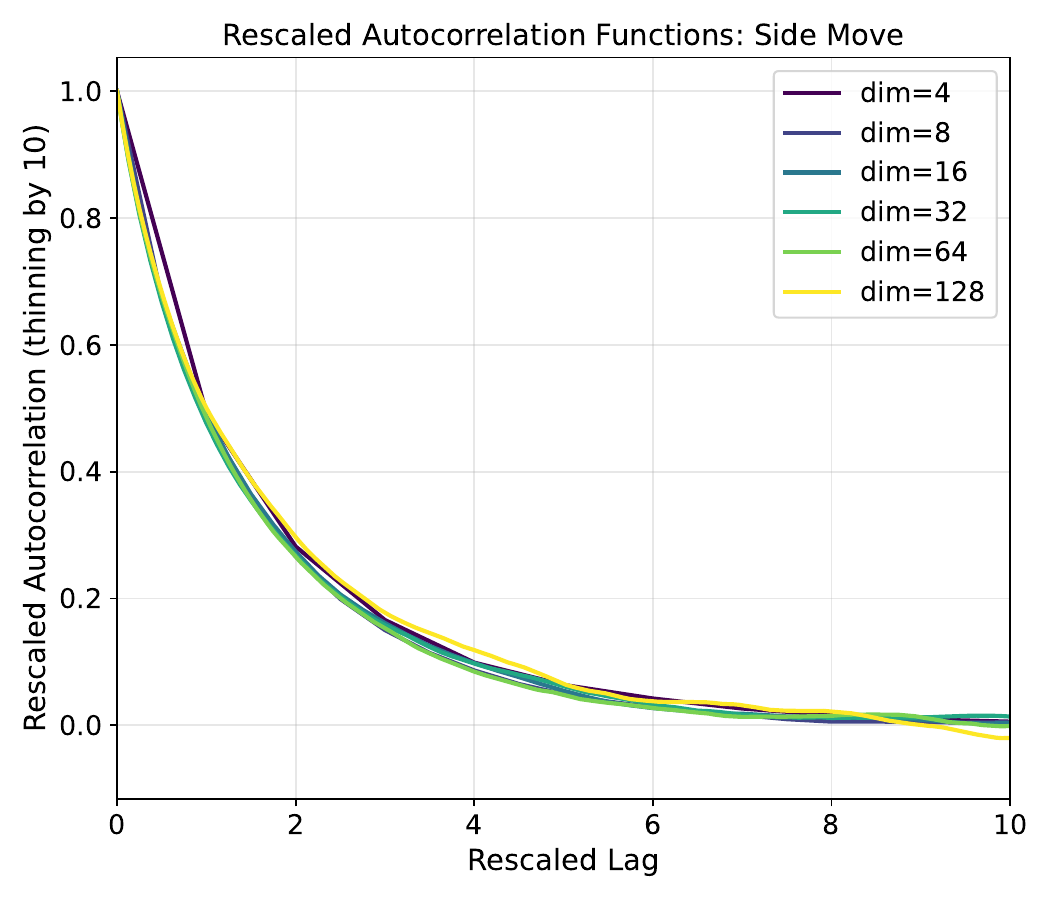}
    \caption{Rescaled autocorrelation functions for sampling anisotropic Gaussian targets with condition number $\kappa = 1000$. Left: stretch move; right: side move. Rescaled lag $= $ original lag$/ $dim$\times 4$.}
    \label{fig:stretch-side-acf}
\end{figure}

In Figure \ref{fig:stretch-side-acf}, we show a scaled version of the autocorrelation functions for the stretch and side moves. We scale the lag by dividing the original lag by the dimension and then multiplying by $4$. With this scaling, the curve represents the unscaled autocorrelation function for $d=4$. We observe that this rescaling produces similar curves for the scaled autocorrelation function, which confirms the high-dimensional linear scaling behavior of both samplers. Notably, the side move leads to faster decay of autocorrelation.

For $d=128$, we report the detailed results to one decimal place in Table \ref{tab:mcmc_performance-Gaussian}.  We observe that the side move achieves approximately $1/2$ of the autocorrelation time of the stretch move. Moreover, with derivative information, samplers become more scalable in high dimensions, as demonstrated by the substantial decrease in autocorrelation time for HMC. We note that when $n=2$, the step size is too large for HMC, causing negligible acceptance rates. However, this large step size works for the affine invariant Hamiltonian walk and side moves. Notably, the Hamiltonian walk move with $n=2$ yields a small autocorrelation time of $1.27$. Compared to HMC, the computational cost is approximately $1/3$, resulting in a speedup of roughly $6.78/1.27 \times 3 \approx 16$ times. If gradient and function evaluations have comparable costs, then the Hamiltonian walk move with $n=2$ achieves $100.01/1.27 \times 1/4 \approx 19.7$ times acceleration compared to the side move. 

Since the Hamiltonian side move employs a side direction (supported on a line) in each iteration, we can reasonably compare it with the stretch and side moves that also move on a line. The Hamiltonian side move leads to shorter autocorrelation time than side move, but it needs more cost (in terms of gradient evaluations) per iteration.
\begin{table}[h]
\centering
\begin{tabular}{l c c c c}
\toprule
 & acceptance & autocorrelation time & func eval & grad eval \\
 & rate &  (thinning by 10) & per iter & per iter \\
\midrule
Stretch move & 0.45 & 204.36 & 1 & 0 \\
Side move & 0.45 & 100.01 & 1 & 0 \\
HMC: $n=10$ & 0.57 & 6.78 & 1 & 11 \\
HMC: $n=2$ & 0.00 & --- & 1 & 3 \\
Hamiltonian walk move: $n=10$ & 0.98 & 1.05 & 1 & 11 \\
Hamiltonian walk move: $n=2$ & 0.61 & 1.27 & 1 & 3 \\
Hamiltonian side move: $n=10$ & 1.00 & 89.82 & 1 & 11 \\
Hamiltonian side move: $n=2$ & 0.98 & 73.23 & 1 & 3 \\
\bottomrule
\end{tabular}
\smallskip
\caption{Performance for anisotropic Gaussian targets with condition number $\kappa = 1000$. Dimension $d=128$. For HMC and affine invariant HMC (Hamiltonian walk and side moves), the total integration time is $T=1$ and $n$ is the number of leapfrog steps.}
\label{tab:mcmc_performance-Gaussian}
\end{table}

\begin{figure}
    \centering
    \includegraphics[width=0.8\linewidth]{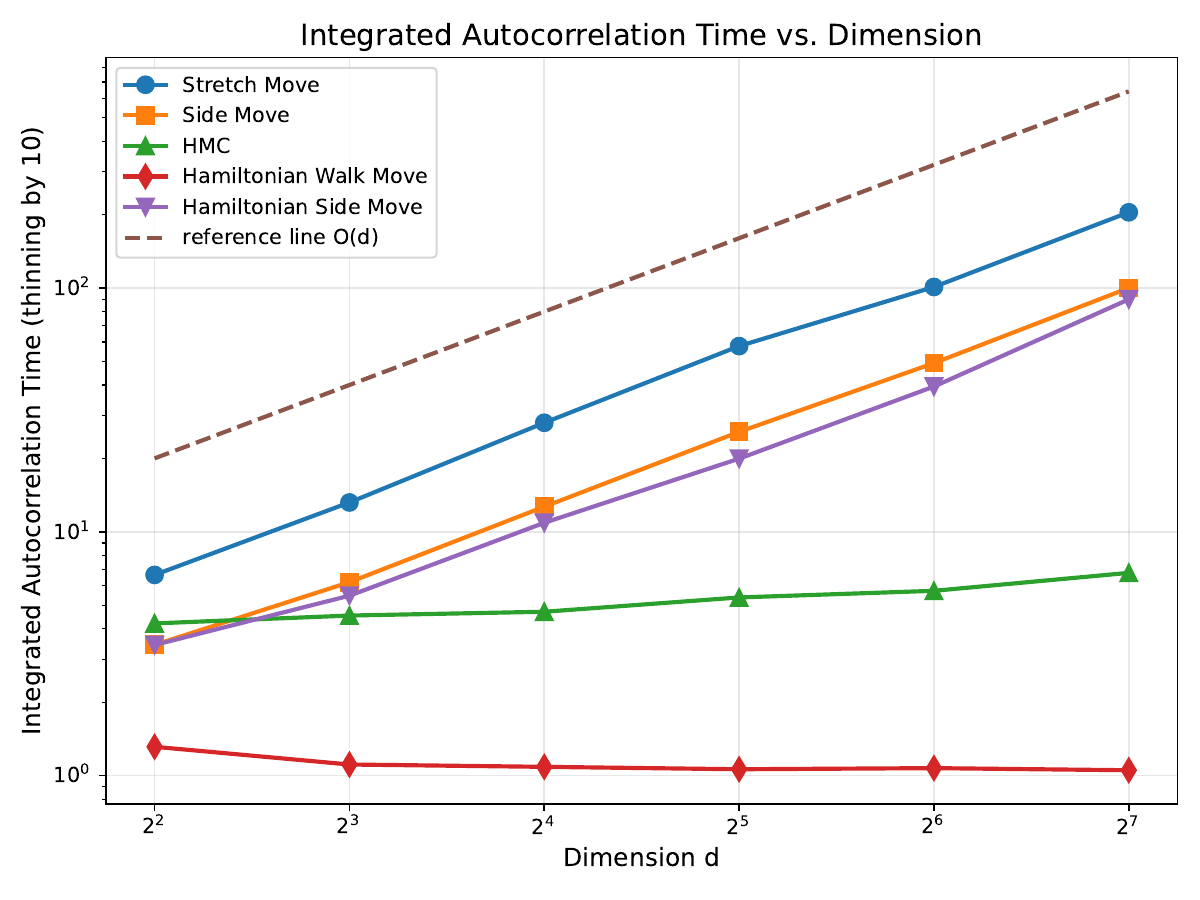}
    \caption{Autocorrelation time (thinning by $10$) versus dimension for anisotropic Gaussian targets with condition number $\kappa = 1000$. For HMC and affine invariant HMC (Hamiltonian walk and side moves), the total integration time is $T=1$, and the number of leapfrog steps is $n=10$ for HMC while $n=2$ for affine invariant HMC.}
    \label{fig:gaussian-act}
\end{figure}

In Figure \ref{fig:gaussian-act}, we further show the autocorrelation time of these samplers. The figure clearly demonstrates the $O(d)$ scaling of stretch and side moves and their Hamiltonian variants. The HMC and Hamiltonian walk move scale much better with dimension, which aligns with the theoretical insights provided by Proposition \ref{thm: scaling-HMC}.

\subsection{Synthetic: Rings} We then consider a ring shaped distribution (motivated by Figure \ref{fig:stretch-side-moves} and discussions therein) with density
\[\pi(\vx) \propto \exp\left(-\frac{(\|\vx\|_2^2 - 1)^2}{l^2}\right)\, , \]
where we choose $d = 50$ and $l = 0.25$. We use the same experimental set-up as in the Gaussian case. In Table \ref{tab:mcmc_performance-ring}, we present the results. Similar phenomenon is observed. The reduction of autocorrelation time of side move compared to stretch move becomes more apparent: the reduction is around $6.8$ times. For this example, standard HMC with $n=10$ performs well, since there is no strong anisotropy among coordinates. In such case, our affine invariant Hamiltonian walk move still leads to a smaller autocorrelation time.

\begin{table}[h]
\centering
\begin{tabular}{l c c c c}
\toprule
 & acceptance & autocorrelation time & func eval & grad eval \\
 & rate & (thinning by 10) & per iter & per iter \\
\midrule
Stretch move & 0.29 & 243.54 & 1 & 0 \\
Side move & 0.45 & 35.54 & 1 & 0 \\
HMC: $n=10$ & 0.69 & 2.08 & 1 & 11 \\
HMC: $n=2$ & 0.00 & --- & 1 & 3 \\
Hamiltonian walk move: $n=10$ & 0.99 & 1.07 & 1 & 11 \\
Hamiltonian walk move: $n=2$ & 0.72 & 1.19 & 1 & 3 \\
Hamiltonian side move: $n=10$ & 1.00 & 35.48 & 1 & 11 \\
Hamiltonian side move: $n=2$ & 0.98 & 30.97 & 1 & 3 \\
\bottomrule
\end{tabular}
\smallskip
\caption{Performance for ring shaped distributions: dimension $d=50$. For HMC and affine invariant HMC (Hamiltonian walk and side moves), the total integration time is $T=1$ and $n$ is the number of leapfrog steps.}
\label{tab:mcmc_performance-ring}
\end{table}

\subsection{Invariant distribution to stochastic PDEs} 
In this illustration, we generate samples from an infinite-dimensional probability measure defined over continuous functions on the unit interval $[0, 1]$; see \cite{goodman2010ensemble}. The measure is formally
\begin{equation}
\exp\left(-\int_0^1 \frac{1}{2}(\partial_x u(x))^2 + V(u(x)){\rm d}x\right),
\end{equation}
with $V$ denoting a double-well potential function:
\[
V(u) = (1 - u^2)^2.
\]
This probability measure is the stationary distribution for the stochastic Allen-Cahn dynamics, which is a stochastic PDE (SPDE):
\begin{equation}
\partial_t u = \partial_{xx}u - V'(u) + \sqrt{2}\,\eta,
\end{equation}
subject to natural boundary conditions at both endpoints $x = 0$ and $x = 1$. Here, $\eta$ denotes space-time white noise. Realizations from this distribution typically exhibit rough, approximately constant profiles near either $1$ or $-1$; thus this is a bimodal distribution. We discretize the derivatives using finite difference with equidistributed points. Denote the total number of points by $d$, which leads to a $d$ dimensional distribution.

\begin{figure}
    \centering
    \includegraphics[width=0.8\linewidth]{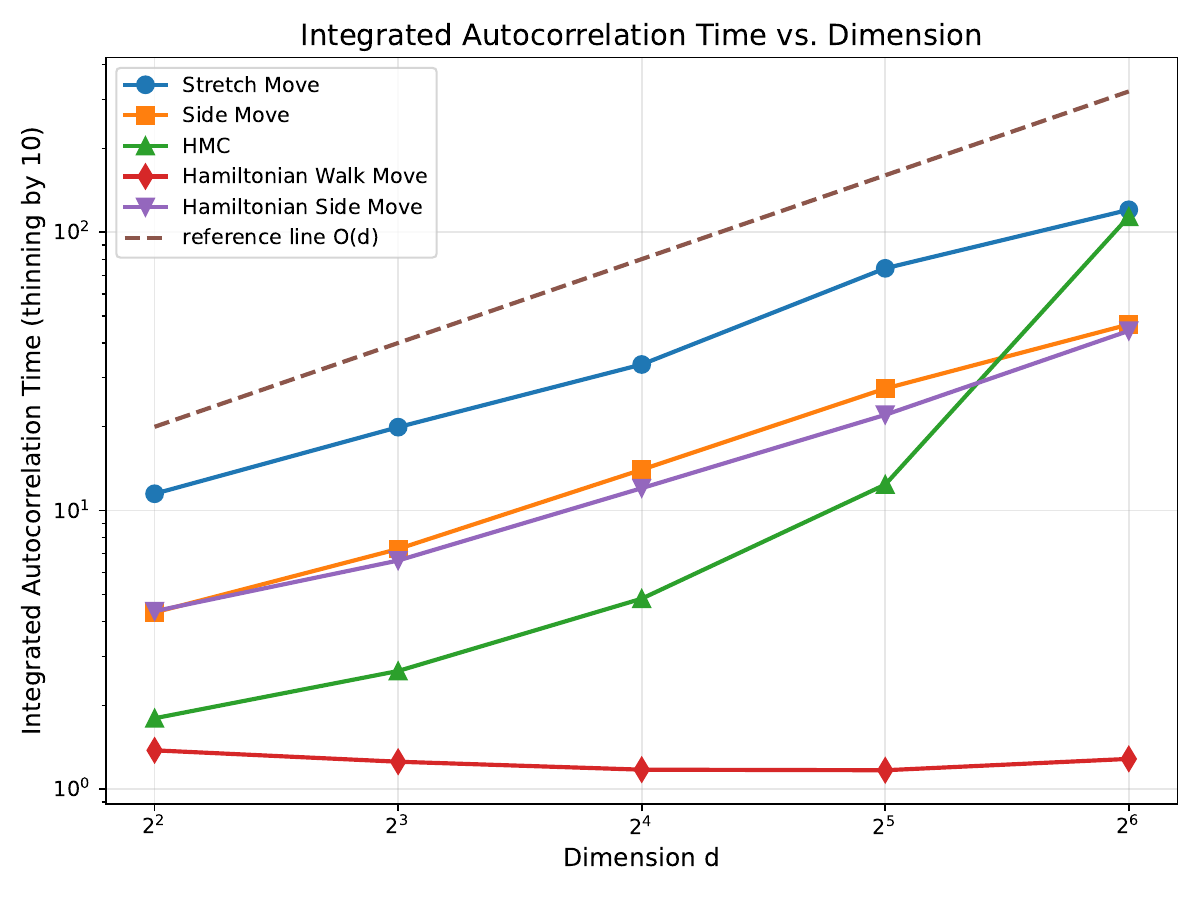}
    \caption{Autocorrelation time (thinning by $10$) versus dimension for the bimodal SPDE targets. For HMC and affine invariant HMC (Hamiltonian walk and side moves), the total integration time is $T=1$, and the number of leapfrog steps is $n=10$ for HMC while $n=2$ for affine invariant HMC.}
    \label{fig:spde-act}
\end{figure}

We adopt the same setup as before. The autocorrelation function is computed for the path integral observable $f(u)=\int_0^1 u(x) {\rm d}x$ discretized using composite trapezoid rules. In Figure \ref{fig:spde-act}, we show the scaling of the autocorrelation time. We observe similar $O(d)$ scaling for stretch and side moves. The HMC scales worse than in the Gaussian example, since the condition number of the SPDE example deteriorates as dimension grows and we use a fixed number of leapfrog steps. For the affine invariant Hamiltonian walk move, we observe a nearly constant autocorrelation time as before, which demonstrates its superior performance in sampling such ill-conditioned distributions. In particular, for $d=64$, Hamiltonian walk move with $n=2$ leapfrog steps leads to $100\times$ reduction in autocorrelation time compared to HMC with $n=10$ leapfrog steps\footnote{In this example with $d=64$, the acceptance rate for HMC is $0.12$, while for the Hamiltonian walk move it is $0.70$.}.

We also report detailed experimental results in Table \ref{tab:mcmc_performance_spde} for $d=128$, a case not included in Figure \ref{fig:spde-act}. For this dimension, standard HMC with $n=10$ fails due to negligible acceptance rates, but the Hamiltonian walk move performs consistently well. 

We note that in this test example, the target distribution is absolutely continuous with respect to a Gaussian measure; accordingly, function space MCMC \cite{cotter2013mcmc} is also expected to achieve dimension-robust convergence. Combining affine invariant samplers at the coarse scale with function space MCMC at the fine scale may further enhance performance \cite{coullon2021ensemble,dunlop2022gradient}.

\begin{table}[h]
\centering
\begin{tabular}{l c c c c}
\toprule
 & acceptance & autocorrelation time & func eval & grad eval \\
 & rate & (thinning by $10$) & per iter & per iter \\
\midrule
Stretch move & 0.44 & 302.13 & 1 & 0 \\
Side move & 0.44 & 139.83 & 1 & 0 \\
HMC: $n=10$ & 0.00 & --- & 1 & 11 \\
HMC: $n=2$ & 0.00 & --- & 1 & 3 \\
Hamiltonian walk move: $n=10$ & 0.98 & 1.12 & 1 & 11 \\
Hamiltonian walk move: $n=2$ & 0.59 & 1.49 & 1 & 3 \\
Hamiltonian side move: $n=10$ & 1.00 & 90.28 & 1 & 11 \\
Hamiltonian side move: $n=2$ & 0.98 & 77.08 & 1 & 3 \\
\bottomrule
\end{tabular}
\smallskip
\caption{Performance for the bimodal distribution which is the invariant distribution to stochastic PDEs: dimension $d=128$. For HMC and affine invariant HMC (Hamiltonian walk and side moves), the total integration time is $T=1$ and $n$ is the number of leapfrog steps.}
\label{tab:mcmc_performance_spde}
\end{table}

\section{Discussions and Conclusions}
\label{sec-Discussions and Conclusions}
In this paper, we propose a derivative-free, affine invariant side move sampler that improves upon the popular stretch move sampler in high dimensions.
We also propose a class of derivative-based, affine invariant HMC samplers, particularly the Hamiltonian walk move, which outperform HMC for sampling from highly anisotropic distributions. We show the dimensional scaling of these samplers for Gaussian targets at the stationary phase, identifying the step size scaling and expected squared jumped distance.

These new affine invariant samplers are shown to be highly efficient on several test examples. Notably, the Hamiltonian walk move sampler achieves a 10 to 100-fold reduction in autocorrelation time compared to vanilla HMC. This highlights the importance of incorporating affine invariance in designing practical samplers.
We anticipate the application of these affine invariant samplers to various scientific problems. 

Algorithmically, adapting the step size and, in the case of affine invariant HMC, the number of leapfrog steps could further improve performance, according to the wide success of the No-U-Turn Sampler (NUTS) \cite{hoffman2014no, carpenter2017stan}; in particular, there has been some recent developments on tuning-free multichain adaptive HMC methods \cite{hoffman2022tuning,hoffman2021adaptive,sountsov2021focusing,sountsov2024running}.  In our experiments, we observe that the Hamiltonian side move leads to smaller autocorrelation time compared to the standard side move. In each iteration, they move in the same direction, suggesting potential for further adaptation of the move length of the derivative-free side move. A recent relevant development is the No-Underrun Sampler (NURS) \cite{bou2025no}, which focuses on adapting move lengths along random line directions without using derivatives.

Theoretically, establishing the ergodicity of these affine invariant ensemble samplers remains an important challenge. To the best of our knowledge, the only work in this direction is \cite{garbuno2020affine} for covariance-preconditioned Langevin dynamics. Still, their analysis applies to continuous dynamics rather than discrete algorithms. Moreover, in addition to the autocorrelation time at the stationary phase, understanding the mixing time or convergence during the burn-in period is also of interest. In this direction, existing results \cite{garbuno2020interacting,carrillo2021wasserstein} primarily focus on Gaussian target distributions in continuous-time covariance-preconditioned Langevin dynamics. We also note that affine invariant algorithms projected onto the Gaussian family (in the sense of variational inference) has been shown to achieve exponential improvement in condition number dependence when the target is a one-dimensional log-concave distribution \cite[Theorem 5.7]{chen2023sampling}.

Finally, beyond affine invariance, there have been developments leveraging diffeomorphism invariant dynamics for sampling, particularly through the Fisher-Rao gradient flow \cite{chen2023sampling}. The numerical approximation of diffeomorphism-invariant dynamics is subtle, and several algorithms can be interpreted as implementations of the Fisher-Rao gradient flow; these include methods based on birth-death processes \cite{lu2019accelerating,lu2023birth}, Kalman methodology and variational inference \cite{huang2022efficient,chen2024efficient,che2025stable}, and ensemble chains \cite{lindsey2022ensemble}. Typically, at most affine invariance is preserved in the numerical implementation and the samplers are often approximate and have bias. It is an interesting direction to integrate these advances to develop exact or approximate samplers with broader invariance properties. 

\vspace{0.2in}
\noindent {\bf Acknowledgments}
YC is grateful to Jonathan Goodman for intellectual discussions on MCMC that inspired this work. YC also thanks Andrew Stuart and Jonathon Weare for helpful communications.
\vspace{0.2in}
\bibliographystyle{plain}
\bibliography{ref}
\appendix
\addtocontents{toc}{\protect\setcounter{tocdepth}{1}}
\section{Pseudocode for All the Affine Invariant Samplers}
\label{appendix-Pseudocode of the Algorithms}
Here we present the pseudocode for the affine invariant samplers used in our experiments:

\begin{itemize}
\item Parallel stretch move (Algorithm \ref{alg-stretch-move})
\item Parallel side move (Algorithm \ref{alg-side-move})
\item Parallel Hamiltonian walk move (Algorithm \ref{alg-H-walk-move})
\item Parallel Hamiltonian side move (Algorithm \ref{alg-H-side-move})
\end{itemize}

The pseudocode describes the iteration step from $m$ to $m+1$. For all the algorithms, we have $N$ walkers in the ensemble and we split the walkers into two groups:
\begin{equation}
    S^{(0)}(m) = \{\mathbf{x}_1(m), ..., \mathbf{x}_{N/2}(m)\}, \quad S^{(1)} = \{\mathbf{x}_{N/2+1}(m),...,\mathbf{x}_N(m)\}\, .
\end{equation}

We use the notation $S^{(-s)}$ to represent the complementary ensemble of $S^{(s)}$ for $s \in \{0,1\}$. Specifically, $S^{(-0)} = S^{(1)}$ and $S^{(-1)} = S^{(0)}$.
\begin{algorithm}
\caption{The parallel stretch move at step $m$ (adapted from \cite{foreman2013emcee})}
\label{alg-stretch-move}
\begin{algorithmic}[1]
\REQUIRE Ensemble $S^{(s)}(m)$ for $s \in \{0, 1\}$, number of walkers $N$, unnormalized probability density $\pi(\cdot)$, parameter $a$
\ENSURE Updated ensemble $S^{(s)}(m+1)$ for $s \in \{0, 1\}$
\FOR{$s \in \{0, 1\}$}
    \FOR{$i = 1 + sN/2, \ldots, N/2 + sN/2$}
        \STATE \textit{// This loop can be done in parallel for all $i$}
        \STATE Draw a walker $\vx_j(m)$ at random from the complementary ensemble $S^{(-s)}(m)$
        \STATE $z \leftarrow Z \sim g$ where $g$ is defined in \eqref{eqn-density-of-stretch}, with the given parameter $a$
        \STATE $\tilde{\vx}_i(m+1) \leftarrow \vx_j(m) + z [\vx_i(m) - \vx_j(m)]$
        \STATE $q \leftarrow z^{d-1} \pi(\tilde{\vx}_i(m+1))/\pi(\vx_i(m))$
        \STATE $r \leftarrow U \sim \mathrm{Unif}[0, 1]$
        \IF{$r \leq q$}
            \STATE $\vx_i(m + 1) \leftarrow \tilde{\vx}_i(m+1)$
        \ELSE
            \STATE $\vx_i(m + 1) \leftarrow \vx_i(m)$
        \ENDIF
    \ENDFOR
\ENDFOR
\end{algorithmic}
\end{algorithm}

\begin{algorithm}
\caption{The parallel side move at step $m$}
\label{alg-side-move}
\begin{algorithmic}[1]
\REQUIRE Ensemble $S^{(s)}(m)$ for $s \in \{0, 1\}$, number of walkers $N$, unnormalized probability density $\pi(\cdot)$, parameter $\sigma$
\ENSURE Updated ensemble $S^{(s)}(m+1)$ for $s \in \{0, 1\}$
\FOR{$s \in \{0, 1\}$}
    \FOR{$i = 1 + sN/2, \ldots, N/2 + sN/2$}
        \STATE \textit{// This loop can be done in parallel for all $i$}
        \STATE Draw two walker $\vx_j(m), \vx_k(m)$ at random from the complementary $S^{(-s)}(m)$
        \STATE $z \leftarrow Z \sim \cN(0,1)$
        \STATE $\tilde{\vx}_i(m+1) \leftarrow \vx_i(m) + \sigma z [\vx_j(m) - \vx_k(m)]$
        \STATE $q \leftarrow \pi(\tilde{\vx}_i(m+1))/\pi(\vx_i(m))$
        \STATE $r \leftarrow U \sim \mathrm{Unif}[0, 1]$
        \IF{$r \leq q$}
            \STATE $\vx_i(m + 1) \leftarrow \tilde{\vx}_i(m+1)$
        \ELSE
            \STATE $\vx_i(m + 1) \leftarrow \vx_i(m)$
        \ENDIF
    \ENDFOR
\ENDFOR
\end{algorithmic}
\end{algorithm}

\begin{algorithm}
\caption{The parallel Hamiltonian walk move at step $m$}
\label{alg-H-walk-move}
\begin{algorithmic}[1]
\REQUIRE Ensemble $S^{(s)}(m)$ for $s \in \{0, 1\}$, number of walkers $N$, unnormalized probability density $\pi(\cdot)$, leapfrog stepsize $h$ and steps $n$
\ENSURE Updated ensemble $S^{(s)}(m+1)$ for $s \in \{0, 1\}$
\FOR{$s \in \{0, 1\}$}
    \FOR{$i = 1 + sN/2, \ldots, N/2 + sN/2$}
        \STATE \textit{// This loop can be done in parallel for all $i$}
        \STATE Draw $\vp_i \sim \cN(0,I_{N/2\times N/2})$
        \STATE Form the centered $B$ for the complementary $S^{(-s)}(m)$ (see \eqref{eqn-centered-ensemble-S1} \eqref{eqn-centered-ensemble-S0})
        \STATE Run $n$ steps leapfrog $(\tilde{\vx}_i(m+1), \tilde{\vp}_i) = \sfP\sfL_{h, B}^n (\vx_i(m), \vp_i)$ as in \eqref{eqn-leapfrog-H-walk-move}
        \STATE $q \leftarrow \exp(-V(\tilde{\vx}_i(m+1))-\frac{1}{2}\tilde{\vp}_i^T\tilde{\vp}_i + V(\vx_i(m)) + \frac{1}{2}\vp_i^T\vp_i)$
        \STATE $r \leftarrow U \sim \mathrm{Unif}[0, 1]$
        \IF{$r \leq q$}
            \STATE $\vx_i(m + 1) \leftarrow \tilde{\vx}_i(m+1)$
        \ELSE
            \STATE $\vx_i(m + 1) \leftarrow \vx_i(m)$
        \ENDIF
    \ENDFOR
\ENDFOR
\end{algorithmic}
\end{algorithm}

\begin{algorithm}
\caption{The parallel Hamiltonian side move at step $m$}
\label{alg-H-side-move}
\begin{algorithmic}[1]
\REQUIRE Ensemble $S^{(s)}(m)$ for $s \in \{0, 1\}$, number of walkers $N$, unnormalized probability density $\pi(\cdot)$, leapfrog stepsize $h$ and steps $n$
\ENSURE Updated ensemble $S^{(s)}(m+1)$ for $s \in \{0, 1\}$
\FOR{$s \in \{0, 1\}$}
    \FOR{$i = 1 + sN/2, \ldots, N/2 + sN/2$}
        \STATE \textit{// This loop can be done in parallel for all $i$}
        \STATE Draw $p_i \sim \cN(0,1)$
        \STATE Draw two walker $\vx_j(m), \vx_k(m)$ at random from the complementary $S^{(-s)}(m)$
        \STATE Run $n$ steps leapfrog $(\tilde{\vx}_i(m+1), \tilde{p}_i) = \sfP\sfL_{h, (\vx_j-\vx_k)/\sqrt{2d}}^n (\vx_i(m), p_i)$ as in \eqref{eqn-H-side-move-leapfrog}
        \STATE $q \leftarrow \exp(-V(\tilde{\vx}_i(m+1))-\frac{1}{2}\tilde{p}_i^T\tilde{p}_i + V(\vx_i(m)) + \frac{1}{2}p_i^Tp_i)$
        \STATE $r \leftarrow U \sim \mathrm{Unif}[0, 1]$
        \ENDFOR
        \IF{$r \leq q$}
            \STATE $\vx_i(m + 1) \leftarrow \tilde{\vx}_i(m+1)$
        \ELSE
            \STATE $\vx_i(m + 1) \leftarrow \vx_i(m)$
        \ENDIF
\ENDFOR
\end{algorithmic}
\end{algorithm}

\section{Proof for the Dimension Scaling of Side and Stretch Moves}
\label{Proof-scaling-side-stretch}
\begin{proof}[Proof for Proposition \ref{prop-gaussian-acceptance}]
For notational convenience, we omit the dependence on $m$.

\subsection{Scaling of side move} For the side move $\tilde{\vx}_i = \vx_i + \sigma (\vx_j-\vx_k) \xi$, we have 
\begin{equation*}
\begin{aligned}
    \log \frac{\pi(\tilde{\vx}_i)}{\pi(\vx_i)} &= - \frac{1}{2}\tilde{\vx}_i^T\tilde{\vx}_i + \frac{1}{2}\vx_i^T\vx_i\\
    & = -\frac{1}{2}\sigma^2 \xi^2 (\vx_j-\vx_k)^T(\vx_j-\vx_k) - \sigma \vx_i \cdot (\vx_j-\vx_k)\xi\, .
\end{aligned}
\end{equation*}
Let $\sigma = \frac{\alpha}{\sqrt{d}}$. By the law of large numbers and central limit theorem, it holds that as $d \to \infty$, 
\begin{equation*}
     \log \frac{\pi(\tilde{\vx}_i)}{\pi(\vx_i)} \overset{\cD}{\longrightarrow} -\alpha^2\xi^2 - \sqrt{2}\alpha \xi z\, ,
\end{equation*}
where $z \sim \cN(0,1)$ and the convergence is in distribution. 

Therefore, the limit of the acceptance probability is
 \[\lim_{d \to \infty} \bE[\min\, \{1, \frac{\pi(\tilde{\vx}_i)}{\pi(\vx_i)} \}] =\bE[\min \{1, \exp(-\alpha^2 \xi^2 - \alpha \sqrt{2} \xi z)\}] \, . \]
This is because the function $t \to \min \{1, \exp(t)\}$ is a bounded continuous function, and we have used the fact that convergence in distribution implies convergence of all bounded continuous observables. The convergence of the expected squared jumped distance follows similarly.

\subsection{Scaling of stretch move} For the stretch move, $\tilde{\vx}_i = \vx_j + Z(\vx_i - \vx_j)$, we have 
\begin{equation*}
\begin{aligned}
     \log \left(Z^{d-1}\frac{\pi(\tilde{\vx}_i)}{\pi(\vx_i)}\right) &= (d-1)\log Z - \frac{1}{2}\tilde{\vx}_i^T\tilde{\vx}_i + \frac{1}{2}\vx_i^T\vx_i \\
     & = (d-1)\log Z - \frac{1}{2}\left((Z^2-1)\vx_i^T\vx_i + (1-Z)^2\vx_j^T\vx_j + 2Z(1-Z)\vx_i^T\vx_j \right)\\
     & = d(\log Z - Z + 1 + \frac{1}{2}(Z-1)^2) - \log Z - \frac{1}{2}d(Z-1)^2 \\
     &\quad - (Z-1)(\frac{Z+1}{2}\vx_i^T\vx_i - Z \vx_i^T\vx_j - d) - \frac{1}{2}(Z-1)^2 \vx_j^T\vx_j\\
     &= d(\log Z - Z + 1 + \frac{1}{2}(Z-1)^2) - \log Z - \frac{1}{2}d(Z-1)^2 \\
     &\quad - (Z-1)(\vx_i^T\vx_i - \vx_i^T\vx_j - d) - \frac{1}{2}(Z-1)^2 \vx_j^T\vx_j\\
     &  \quad - \frac{(Z-1)^2}{2}\vx_i^T\vx_i + (Z-1)^2 \vx_i^T\vx_j\, ,
     \end{aligned}
\end{equation*}
where we organize the terms in a way that is convenient to apply limit theorems later.

Here, the $Z$ has the density
\begin{equation*}
    g(z) \propto \begin{cases}
\frac{1}{\sqrt{z}} & \text{if } z \in \left[\frac{1}{a}, a\right] \\
0 & \text{otherwise}\, ,
\end{cases}
\end{equation*}
which implies that we can write $Z = \left(\frac{(\sqrt{a}-\sqrt{a^{-1}})U + \sqrt{a}+\sqrt{a^{-1}}}{2}\right)^2$ for $U \sim \mathrm{Unif}[-1,1]$. Using the formula, we have $\sqrt{d}(Z-1) \overset{a.s.}{\longrightarrow} \beta U$ for $a = 1+\frac{\beta}{\sqrt{d}}$. 

By direct calculation, as $d \to \infty$, we have
\begin{itemize}
    \item $d(\log Z - Z + 1 + \frac{1}{2}(Z-1)^2) \overset{a.s.}{\longrightarrow} 0 .$
    \item $\log Z \overset{a.s.}{\longrightarrow} 0 .$
\end{itemize}
These show that several parts in the formula of $\log \left(Z^{d-1}\frac{\pi(\tilde{\vx}_i)}{\pi(\vx_i)}\right)$ will converge to zero almost surely. For the remaining part in the formula of $\log \left(Z^{d-1}\frac{\pi(\tilde{\vx}_i)}{\pi(\vx_i)}\right)$, we note that the random vector \[(\sqrt{d}(Z-1), \frac{1}{\sqrt{d}}(\vx^T_i\vx_i - d), \frac{1}{\sqrt{d}}\vx_i^T\vx_j, \frac{1}{d}\vx_i^T\vx_i, \frac{1}{d}\vx_i^T\vx_j)\]
converges in distribution to $(\beta U, z_1, z_2, \delta_{1}, \delta_0)$ (for convergence in distribution in the product space, see \cite[Section 2]{billingsley2013convergence}). Here $U \sim \mathrm{Unif}[-1,1], z_1 \sim \cN(0,3), z_2 \sim \cN(0,1)$ are independent, and $\delta_x$ is a Diracs point pass at $x$. The convergence is obtained using the law of large numbers and the central limit theorem as well as the fact that $Z$ and $\vx_i, \vx_j$ are independent. We also used the fact that convergence almost surely implies convergence in distribution.

Then, the remaining part in $\log \left(Z^{d-1}\frac{\pi(\tilde{\vx}_i)}{\pi(\vx_i)}\right)$ can be seen as a continuous function of the above random vector. Using the property of convergence in distribution, we get that
\begin{equation}
    \log \left(Z^{d-1}\frac{\pi(\tilde{\vx}_i)}{\pi(\vx_i)}\right) \overset{\cD}{\longrightarrow} -\frac{3}{2}\beta^2U^2 - \sqrt{3}\beta U z\, ,
\end{equation}
where $U \sim \mathrm{Unif}[-1,1]$ is independent of $z \sim \cN(0,1)$. This then implies the convergence of the acceptance probability and the squared jumped distance as desired.
\end{proof}
\section{Proof for the Dimension Scaling of Affine Invariant HMC}
\label{Proof-scaling-HMC}
\begin{proof}[Proof for Proposition \ref{thm: scaling-HMC}]
    We focus on one particle $\vx_i$. For simplicity of notation, we omit the subindex $i$ when there is no confusion. We analyze one iteration in the affine invariant HMC: we use superscript $\vx^{(l)}$ to denote the result from the leapfrog scheme, and we use $B$ for a general preconditioner which can either be the one in the Hamiltonian walk move or the side move. By definition,
    \begin{equation*}
        \begin{aligned}
            \vp^{(l+1/2)} &= \vp^{(l)} - \frac{h}{2}B^T\nabla V(\vx)\\
            \vx^{(l+1)} & = \vx^{(l)} + hB\vp^{(l+1/2)} \\
            \vp^{(l+1)} &= \vp^{(l+1/2)} - \frac{h}{2}B^T\nabla V(\vx^{(l+1)})\, .
        \end{aligned}
    \end{equation*}
The above leads to the iteration
\begin{equation*}
    \begin{bmatrix}
B^T \vx^{(l+1)} \\
\vp^{(l+1)}
\end{bmatrix}
=
\begin{bmatrix}
I - \frac{h^2}{2}B^TB & hB^TB \\
-h (I - \frac{h^2}{4}B^TB) & I - \frac{h^2}{2}B^TB
\end{bmatrix}
\begin{bmatrix}
B^T\vx^{(l)} \\
\vp^{(l)}
\end{bmatrix},
\end{equation*}
where $B^T \vx^{(l)}$ and $\vp^{(l)}$ are of the same dimension; we denote the dimension by $D$. We will use the notation $\vq^{(l)} = B^T\vx^{(l)}$. 

\subsection{A simple formula for the acceptance probability} First, we show that the acceptance probability can be computed using the coordinates $\vq^{(n)}$ and $\vp^{(n)}$. To do so, note that we have the relation 
\[\vx^{(n)}-\vx^{(0)} = B(B^TB)^{+}(\vq^{(n)}-\vq^{(0)})\, , \]
where $(B^TB)^{+}$ is the Moore–Penrose inverse of the matrix $B^TB \in \bR^{D\times D}$.

This implies that 
\begin{equation*}
    \begin{aligned}
        \frac{|\vx^{(n)}|^2_2}{2} -\frac{|\vx^{(0)}|^2_2}{2}  &= \frac{1}{2}|\vx^{(0)}+B(B^TB)^{+}(\vq^{(n)}-\vq^{(0)})|_2^2 - \frac{1}{2}|\vx^{(0)}|^2_2\\
        & = (\vx^{(0)})^T B(B^TB)^{+}(\vq^{(n)}-\vq^{(0)}) + \frac{1}{2}(\vq^{(n)}-\vq^{(0)})^T(B^TB)^{+}(\vq^{(n)}-\vq^{(0)})\\
        & = (\vq^{(0)})^T(B^TB)^{+}(\vq^{(n)}-\vq^{(0)}) + \frac{1}{2}(\vq^{(n)}-\vq^{(0)})^T(B^TB)^{+}(\vq^{(n)}-\vq^{(0)})\\
        & = \frac{1}{2}(\vq^{(n)})^T(B^TB)^{+}\vq^{(n)} - \frac{1}{2}(\vq^{(0)})^T(B^TB)^{+}\vq^{(0)}\, .
    \end{aligned}
\end{equation*}
Thus the acceptance probability can be written in the $\vq^{(n)}$ and $\vp^{(n)}$ coordinates:
\[\min\, \{1, \exp(-\frac{(\vq^{(n)})^T(B^TB)^{+}\vq^{(n)}+ (\vp^{(n)})^T\vp^{(n)}}{2} + \frac{(\vq^{(0)})^T(B^TB)^{+}\vq^{(0)} + (\vp^{(0)})^T\vp^{(0)}}{2})\}\, .  \]

Let $B^TB = U\Sigma U^T \in \bR^{D\times D}$ be the eigenvalue decomposition where $U$ is an orthogonal matrix and $\Sigma = \mathrm{diag}(\lambda_1^2,...,\lambda_r^2, \lambda_{r+1}^2, ..., \lambda_D^2)$. Here we assume the rank of $B^TB$ to be $r$ and $\lambda_1 \geq \lambda_2 \geq ... \geq \lambda_r > 0 = \lambda_{r+1} = \cdots = \lambda_D$.
Let $\bar{\vq}^{(l)} = U^T \vq^{(l)}$ and $\bar{\vp}^{(l)} = U^T \vp^{(l)}$. Then the acceptance probability is 

    \[\min\, \{1, \exp(-\frac{(\bar{\vq}^{(n)})^T\Sigma^{+}\bar{\vq}^{(n)}+ (\bar{\vp}^{(n)})^T\bar{\vp}^{(n)}}{2} + \frac{(\bar{\vq}^{(0)})^T\Sigma^{+}\bar{\vq}^{(0)} + (\bar{\vp}^{(0)})^T\bar{\vp}^{(0)}}{2})\}\, .  \]
Moreover, we have the iteration
    \begin{equation*}
    \begin{bmatrix}
\bar{\vq}^{(l+1)} \\
\bar{\vp}^{(l+1)}
\end{bmatrix}
=
\begin{bmatrix}
I - \frac{h^2}{2}\Sigma & h\Sigma \\
-h (I - \frac{h^2}{4}\Sigma) & I - \frac{h^2}{2}\Sigma
\end{bmatrix}
\begin{bmatrix}
\bar{\vq}^{(l)} \\
\bar{\vp}^{(l)}
\end{bmatrix},
\end{equation*}
Denote $\bar{\vq}^{(l)} = (\bar{q}_1^{(l)},...,\bar{q}_D^{(l)})$ and $\bar{\vp}^{(l)} = (\bar{p}_1^{(l)},...,\bar{p}_D^{(l)})$. For the $i$-th coordinate, we have
\begin{equation*}
    \begin{bmatrix}
\bar{q}^{(l+1)}_i \\
\bar{p}^{(l+1)}_i
\end{bmatrix}
=
\begin{bmatrix}
1 - \frac{h^2}{2}\lambda_i^2 & h\lambda_i^2 \\
-h (1 - \frac{h^2}{4}\lambda_i^2) & 1 - \frac{h^2}{2}\lambda_i^2
\end{bmatrix}
\begin{bmatrix}
\bar{q}^{(l)}_i \\
\bar{p}^{(l)}_i
\end{bmatrix}\, .
\end{equation*}
We can solve the above system explicitly.
First, for $r+1 \leq i \leq D$, we get $\bar{q}_i^{(l+1)} = \bar{q}_i^{(l)}$ and $\bar{p}^{(l+1)}_i = -h \bar{q}_i^{(l)} + \bar{p}^{(l)}_i$. Recall $\bar{\vq}^{(0)} = U^TB^T\vx^{(0)} \sim \cN(0, U^TB^TBU) = \cN(0,\Sigma)$. This means that $\bar{q}_i^{(0)} = 0$ for $r+1 \leq i \leq D$. Thus $\bar{q}_i^{(l)} = 0$ for all $l$ and $\bar{p}^{(l+1)}_i = \bar{p}^{(l)}_i$ for all $l$. Therefore the term that depends on $r+1 \leq i \leq D$ in the acceptance probability satisfies
\[ -\frac{|\bar{p}^{(n)}_i|^2}{2} + \frac{|\bar{p}^{(0)}_i|^2}{2} = 0\, . \]

For $1\leq i \leq r$, let us assume $h \leq 2/\lambda_i$. The eigenvalues of the matrix are 
\[\mu = 1-\frac{h^2}{2}\lambda_i^2 \pm \mathsf{j} \left(h\lambda_i \sqrt{1-\frac{1}{4}h^2\lambda_i^2}\right) \]
where $\mathsf{j}$ is the imaginary number. Denote $\phi_i \in [0,\pi]$ such that $\cos \phi_i = 1 - \frac{h^2\lambda_i^2}{2}, \sin \phi_i = h\lambda_i \sqrt{1-\frac{1}{4}h^2\lambda_i^2}$. We get
\begin{equation}
\label{eqn-discrete-HD}
    \begin{bmatrix}
\bar{q}^{(l+1)}_i \\
\bar{p}^{(l+1)}_i
\end{bmatrix}
=
\begin{bmatrix}
\cos \phi_i & \hat{\lambda}_i \sin\phi_i \\
-\frac{1}{\hat{\lambda}_i}\sin \phi_i & \cos \phi_i
\end{bmatrix}
\begin{bmatrix}
\bar{q}^{(l)}_i \\
\bar{p}^{(l)}_i
\end{bmatrix}\, ,
\end{equation}
where $\hat{\lambda}_i = \frac{\lambda_i}{\sqrt{1-h^2\lambda_i^2/4}}$. Iterating the equation leads to the solution
\begin{equation*}
    \begin{bmatrix}
\bar{q}^{(n)}_i \\
\bar{p}^{(n)}_i
\end{bmatrix}
=
\begin{bmatrix}
\cos (n\phi_i) & \hat{\lambda}_i \sin(n\phi_i) \\
-\frac{1}{\hat{\lambda}_i}\sin (n\phi_i) & \cos (n\phi_i)
\end{bmatrix}
\begin{bmatrix}
\bar{q}^{(0)}_i \\
\bar{p}^{(0)}_i
\end{bmatrix}\, .
\end{equation*}
On the other hand, consider the Hamiltonian dynamics
\[ \frac{{\rm d} \bar{q}(t) }{{\rm d} t } = \hat{\lambda}_i^2 \bar{p}(t), \quad \frac{{\rm d} \bar{p}(t) }{{\rm d} t } = - \bar{q}(t)\, , \]
which has an explicit solution
\begin{equation}
\label{eqn-continuous-HD}
    \begin{bmatrix}
\bar{q}(t) \\
\bar{p}(t)
\end{bmatrix}
=
\begin{bmatrix}
\cos (\hat{\lambda}_i t) & \hat{\lambda}_i \sin(\hat{\lambda}_i t) \\
-\frac{1}{\hat{\lambda}_i}\sin (\hat{\lambda}_i t) & \cos (\hat{\lambda}_i t)
\end{bmatrix}
\begin{bmatrix}
\bar{q}(0)\\
\bar{p}(0)
\end{bmatrix}\, .
\end{equation}
By taking the same initial conditions for \eqref{eqn-discrete-HD} and \eqref{eqn-continuous-HD}, and $\hat{\lambda}_i t = n \phi_i$, we get the same solution for the discrete and continuous systems. 

Since \eqref{eqn-continuous-HD} is the exact solution to a Hamiltonian dynamics, we have conservation of several quantities along the dynamics. In particular, for a specific energy, we get
\[ \frac{|\bar{q}(t)|^2}{2\hat{\lambda}_i^2} + \frac{|\bar{p}(t)|^2}{2} = \frac{|\bar{q}(0)|^2}{2\hat{\lambda}_i^2} + \frac{|\bar{p}(0)|^2}{2} \, . \]
Due to the equivalence mentioned above, we thus get
\[ \frac{|\bar{q}^{(n)}_i|^2}{2\hat{\lambda}_i^2} + \frac{|\bar{p}^{(n)}_i|^2}{2} = \frac{|\bar{q}^{(0)}_i|^2}{2\hat{\lambda}_i^2} + \frac{|\bar{p}^{(0)}_i|^2}{2} \, . \]
Therefore, the $i$-th term (for $1\leq i \leq r$) in the acceptance probability satisfies
\begin{equation}
\begin{aligned}
     &-\frac{|\bar{q}^{(n)}_i|^2}{2{\lambda}_i^2} - \frac{|\bar{p}^{(n)}_i|^2}{2} + \frac{|\bar{q}^{(0)}_i|^2}{2{\lambda}_i^2} + \frac{|\bar{p}^{(0)}_i|^2}{2}\\
     =& -\frac{|\bar{q}^{(n)}_i|^2}{2{\lambda}_i^2} + \frac{|\bar{q}^{(n)}_i|^2}{2\hat{\lambda}_i^2} + \frac{|\bar{q}^{(0)}_i|^2}{2{\lambda}_i^2} - \frac{|\bar{q}^{(0)}_i|^2}{2\hat{\lambda}_i^2}\\
     =& \frac{h^2\lambda_i^4}{8}(-|\bar{q}^{(n)}_i|^2 + |\bar{q}^{(0)}_i|^2)\, .
\end{aligned} 
\end{equation}
Then the overall acceptance rate is accordingly
\begin{equation}
\label{eqn-appendix-acceptance-prob}
    \begin{aligned}
    &\min\, \{1, \exp(-\frac{(\bar{\vq}^{(n)})^T\Sigma^{+}\bar{\vq}^{(n)}+ (\bar{\vp}^{(n)})^T\bar{\vp}^{(n)}}{2} + \frac{(\bar{\vq}^{(0)})^T\Sigma^{+}\bar{\vq}^{(0)} + (\bar{\vp}^{(0)})^T\bar{\vp}^{(0)}}{2})\}\\
    = &\min\, \{1, \exp(\frac{h^2}{8}\sum_{i=1}^r \lambda_i^4 (|\bar{q}^{(0)}_i|^2 - |\bar{q}^{(n)}_i|^2))\}\, .
    \end{aligned}
\end{equation}
This formula is similar to the one derived in \cite{apers2024hamiltonian} for standard Hamiltonian dynamics. We adapt the strategy to derive the formula for the above preconditioned Hamiltonian dynamics. Here, the additional $B$ requires us to discuss the rank of the matrix $B^TB$.

\subsection{Large $d$ limit of acceptance  probability for the Hamiltonian walk move}
In this case, we have $B\in \mathbb{R}^{d\times N/2}$ and $D = N/2 \geq d$. The rank of $B^TB$ is thus $r =d$.

Now, for the $i$-th term, since $\bar{q}^{(n)}_i = \cos(n\phi_i)\bar{q}^{(0)}_i + \hat{\lambda}_i \sin(n\phi_i)\bar{p}^{(0)}_i$,
we have
\begin{equation}
\label{eqn-appendix-q0-qn}
    |\bar{q}^{(0)}_i|^2 - |\bar{q}^{(n)}_i|^2 = (1-\cos^2(n\phi_i))|\bar{q}^{(0)}_i|^2 - \hat{\lambda}_i^2 \sin^2(n\phi_i)|\bar{p}^{(0)}_i|^2 - 2\hat{\lambda}_i\cos(n\phi_i)\sin(n\phi_i)\bar{q}^{(0)}_i\bar{p}^{(0)}_i\, .
\end{equation}
Recall our assumption: $\vx^{(0)}, \vp^{(0)}$ are independent multivariate normal distributions. This implies that $\bar{q}^{(0)}_i \sim \cN(0, \lambda_i^2)$. By a direct and tedious calculation, we get that for $h \leq \min_{1\leq i\leq d}\frac{2}{\lambda_i}$,
\begin{equation}
\begin{aligned}
    &\bE[\lambda_i^4(|\bar{q}^{(0)}_i|^2 - |\bar{q}^{(n)}_i|^2)] = -\frac{h^2\lambda_i^8}{4}\sin^2(n\phi_i)\frac{1}{1-h^2\lambda_i^2/4}\, ,\\
    & \mathrm{Var}[\lambda_i^4(|\bar{q}^{(0)}_i|^2 - |\bar{q}^{(n)}_i|^2)] = \frac{4\lambda_i^{12}}{1- h^2\lambda_i^2/4}\sin^2(n\phi_i) + O(h^4) \, .
\end{aligned}
\end{equation}
Therefore
\begin{equation}
\label{eqn-appendix-expectation}
    \bE[\frac{h^2}{8}\sum_{i=1}^d \lambda_i^4 (|\bar{q}^{(0)}_i|^2 - |\bar{q}^{(n)}_i|^2)] = -\frac{h^4}{32} \sum_{i=1}^d \frac{\lambda_i^8}{1-h^2\lambda_i^2/4} \sin^2(n\phi_i)\, ,
\end{equation}
and 
\begin{equation}
\label{eqn-appendix-var}
    \mathrm{Var}[\frac{h^2}{8}\sum_{i=1}^d \lambda_i^4 (|\bar{q}^{(0)}_i|^2 - |\bar{q}^{(n)}_i|^2)] = (\frac{h^4}{16} + O(h^8)) \sum_{i=1}^d\frac{\lambda_i^{12}}{1- h^2\lambda_i^2/4}\sin^2(n\phi_i) \, .
\end{equation}
We also note that $\cos \phi_i = 1 - \frac{h^2\lambda_i^2}{2}$, so 
\[ \phi_i = \mathrm{arccos}(1 - \frac{h^2\lambda_i^2}{2}) = h \lambda_i + \frac{(h\lambda_i)^3}{24} + O(h^4)\, , \]
which implies that if we take $n = \lceil\frac{T}{h}\rceil$ with $T$ fixed, then
\[ \cos(n\phi_i) = \cos(\lambda_i T + O(h^2)), \quad \sin(n\phi_i) = \sin(\lambda_i T + O(h^2)) \, .  \]

We are now in the position to derive the $d\to \infty$ limit of the acceptance probability. By writing $N$ as a function of $d$ and assuming $\lim_{d\to \infty} \frac{d}{N(d)/2} = \rho < 1$, we get that the spectral measure 
\[\mu_d = \frac{1}{d}\sum_{i=1}^d \delta_{\lambda_i^2}\] 
of the random matrix $B^TB$ will converge almost surely to 
\[\mathrm{d}\nu_{\rho}(\lambda) = \frac{1}{2\pi \rho \lambda} \sqrt{(c-\lambda)(\lambda-b)}\chi_{[b,c]}(\lambda)\mathrm{d}\lambda
\]
where $b = (1-\sqrt{\rho})^2, c = (1+\sqrt{\rho})^2$ and $\chi_{[b,c]}$ is the characteristic function of $[b,c]$. This is due to the Mar\v{c}enko--Pastur law. These limiting eigenvalues all belong to $[b,c]$ which is independent of $d$. Taking $h = \alpha d^{-1/4}, n = \lceil\frac{T}{h}\rceil$ and using the central limit theorem with \eqref{eqn-appendix-expectation} and \eqref{eqn-appendix-var}, we get
\begin{equation}
    \frac{h^2}{8}\sum_{i=1}^d \lambda_i^4 (|\bar{q}^{(0)}_i|^2 - |\bar{q}^{(n)}_i|^2) \overset{\cD}{\longrightarrow} \cN(\alpha^4 \mu_{\rho}, \alpha^4\sigma_{\rho})
\end{equation}
where
\[\mu_{\rho} = -\frac{1}{32}\int \lambda^{4} \sin^2(\sqrt{\lambda} T) {\rm d}\nu_{\rho}(\lambda)\]
and
\[\sigma_\rho = \frac{1}{16}\int \lambda^{6} \sin^2(\sqrt{\lambda} T) {\rm d}\nu_{\rho}(\lambda)\, .\]
Thus, the acceptance probability converges to 
\[ \bE[\min\{1, \exp(\cN(\alpha^4 \mu_{\rho}, \alpha^4\sigma_{\rho}))\}]\, . \]

\subsection{Large $d$ limit of acceptance  probability for the Hamiltonian side move} In this case, we have $D=1$ and $B \in \bR^{d\times 1}$. Indeed, $B^TB = \frac{1}{2d}(\vx_j - \vx_k)^T(\vx_j - \vx_k) \to 1$ almost surely as $d \to \infty$. So the eigenvalue $\lambda_1^2$ of $B^TB$ converges to $1$, and the rank of $B^TB$ is $r = 1$.

Based on \eqref{eqn-appendix-acceptance-prob}, the acceptance rate has the following formula
\[\min\, \{1, \exp(\frac{h^2}{8} \lambda^4_1 (|\bar{q}^{(0)}_1|^2 - |\bar{q}^{(n)}_1|^2)\}\, ,\]
where $\bar{q}^{(0)}_1 \sim \cN(0,\lambda^2_1)$, $\bar{q}^{(n)}_1 = \cos(n\phi)\bar{q}^{(0)}_1 + \hat{\lambda}_1 \sin(n\phi)\bar{p}^{(0)}_1$ with $\bar{p}^{(0)}_1 \sim \cN(0,1)$ and $\hat{\lambda}_1 = \frac{\lambda_1}{\sqrt{1-h^2\lambda^2_1/4}}$.

Taking $h = \alpha < 2$, using the formula \eqref{eqn-appendix-q0-qn}, we then get as $d \to \infty$,
\begin{equation}
    \frac{h^2}{8} \lambda^4_1 (|\bar{q}^{(0)}_1|^2 - |\bar{q}^{(n)}_1|^2) \overset{a.s.}{\longrightarrow} \frac{\alpha^2}{8}\left(\sin^2(n\phi)(z_1^2 - z_2^2) - \sin(2n\phi)z_1z_2\right)
\end{equation}
where $z_1 \sim \cN(0,1), z_2\sim \cN(0,\frac{1}{1-\alpha^2/4})$ are independent, and $\phi \in [0,\pi]$ such that $\cos \phi = 1 - \frac{\alpha^2}{2}$.
The acceptance probability converges to 
\[ \bE[\min\{1, \exp(\frac{\alpha^2}{8}\left(\sin^2(n\phi)(z_1^2 - z_2^2) - \sin(2n\phi)z_1z_2\right))\}]\, . \]

\subsection{Expected squared jumped distance} Next, we derive the asymptotic limit of the expected squared jumped distance. We consider the general setting and then take specific values of $B$ to obtain results for Hamiltonian walk and side moves. 

The squared jumped distance for each iteration is 
\[\|\vx^{(n)}-\vx^{(0)}\|_2^2 = \|B(B^TB)^{+}(\vq^{(n)}-\vq^{(0)})\|_2^2 = \sum_{i=1}^r \frac{1}{\lambda_i^2} |\bar{q}^{(n)}_i - \bar{q}^{(0)}_i|^2\, , \]
where we used the fact that $\|B(B^TB)^{+}(\vq^{(n)}-\vq^{(0)})\|_2^2 = (\vq^{(n)}-\vq^{(0)})^T(B^TB)^{+}(\vq^{(n)}-\vq^{(0)})$ and $B^TB = U\Sigma U^T \in \bR^{D\times D}$, $\bar{\vq}^{(l)} = U^T \vq^{(l)}$ with $\bar{\vq}^{(l)} = (\bar{q}_1^{(l)},...,\bar{q}_D^{(l)})$. We note that $r$ is the rank of $B^TB$. More specifically, $r=d$ in the Hamiltonian walk move and $r=1$ in the Hamiltonian side move.

The expected squared jumped distance in each iteration then admits the formula
\[\bE[\left(\sum_{i=1}^r \frac{1}{\lambda_i^2} |\bar{q}^{(n)}_i - \bar{q}^{(0)}_i|^2\right) \min\, \{1, \exp(\frac{h^2}{8}\sum_{i=1}^r \lambda_i^4 (|\bar{q}^{(0)}_i|^2 - |\bar{q}^{(n)}_i|^2))\}] \, . \]
We recall that $\bar{q}^{(0)}_i \sim \cN(0, \lambda_i^2)$ are independent Gaussian random variables for each $i$, and 
\[\bar{q}^{(n)}_i = \cos(n\phi_i)\bar{q}^{(0)}_i + \hat{\lambda}_i \sin(n\phi_i)\bar{p}^{(0)}_i\]
where $\bar{p}^{(0)}_i$ are independent standard normal distributions for each $i$.

First, we consider the Hamiltonian walk move. Here $r=d$. For each $i$, 
\[\frac{|\bar{q}^{(n)}_i - \bar{q}^{(0)}_i|^2}{\lambda_i^2}  = \frac{(\cos(n\phi_i) - 1)^2}{\lambda_i^2}(\bar{q}^{(0)}_i)^2 + \frac{\hat{\lambda}_i^2}{\lambda_i^2}\sin^2(n\phi_i)(\bar{p}^{(0)}_i)^2 + \frac{2\hat{\lambda}_i}{\lambda_i^2}(\cos(n\phi_i)-1)\sin(n\phi_i) \bar{q}^{(0)}_i \bar{p}^{(0)}_i \, .\]
By writing $N$ as a function of $d$ and assuming $\lim_{d\to \infty} \frac{d}{N(d)/2} = \rho < 1$, and using law of large numbers, we get
\begin{equation*}
\begin{aligned}
    \lim_{N\to\infty}\frac{1}{d}\sum_{i=1}^{d} \frac{1}{\lambda_i^2} |\bar{q}^{(n)}_i - \bar{q}^{(0)}_i|^2 &= \int \left((\cos(\sqrt{\lambda} T) - 1)^2 + \sin^2(\sqrt{\lambda} T) \right) {\rm d}\nu_{\rho}(\lambda)\\ 
    &= \int 4\sin^2(\frac{\sqrt{\lambda} T}{2}){\rm d}\nu_{\rho}(\lambda)\, .
\end{aligned}
\end{equation*}
Second, for the Hamiltonian side move, we have as $d\to \infty$,
\[\frac{|\bar{q}^{(n)}_1 - \bar{q}^{(0)}_1|^2}{\lambda_1^2} \overset{a.s.}{\longrightarrow} (\cos(n\phi)-1)^2z_1^2 + \sin^2(n\phi)z_2^2 + 2(\cos(n\phi)-1)\sin(n\phi)z_1z_2\, , \]
where again, $z_1 \sim \cN(0,1), z_2\sim \cN(0,\frac{1}{1-\alpha^2/4})$ are independent, and $\phi \in [0,\pi]$ such that $\cos \phi = 1 - \frac{\alpha^2}{2}$. Thus the expected squared jumped distance is
\[\bE[Q(z_1,z_2, n, \alpha)\min\{1, \exp(\frac{\alpha^2}{8}\left(\sin^2(n\phi)(z_1^2 - z_2^2) - \sin(2n\phi)z_1z_2\right))\} ] \]
where 
\[Q(z_1,z_2, n, \alpha) = (\cos(n\phi)-1)^2z_1^2 + \sin^2(n\phi)z_2^2 + 2(\cos(n\phi)-1)\sin(n\phi)z_1z_2\, .\]
\end{proof}

\end{document}